\documentclass[fleqn,usenatbib]{rasti}

\usepackage{newtxtext,newtxmath}

\usepackage[T1]{fontenc}

\DeclareRobustCommand{\VAN}[3]{#2}
\let\VANthebibliography\thebibliography
\def\thebibliography{\DeclareRobustCommand{\VAN}[3]{##3}\VANthebibliography}

\usepackage{graphicx}	
\usepackage{amsmath}	
\usepackage{indentfirst}
\usepackage{caption}
\usepackage{threeparttable} 
\usepackage{booktabs} 
\usepackage{multirow}

\title[1D CNN Retrieval of Reflection Spectra]{Towards the Habitable Worlds Observatory: Retrieval of Reflection Spectra from Evolving Earth Analogues using 1D CNNs}

\author[S. G. A. Barbosa et al.]{
Sarah G. A. Barbosa,$^{1}$\thanks{E-mail: sarah@fisica.ufc.br (SGAB)}
Raissa Estrela,$^{2}$
Paulo C. F. da Silva Filho,$^{1}$
Lorenzo V. Mugnai, $^{3}$\newauthor
\,and  Daniel B. de Freitas $^{1}$  
\\
$^{1}$Department of Physics, Federal University of Ceará, Campus do Pici, Fortaleza, CE 60455-900, Brazil\\
$^{2}$Jet Propulsion Laboratory, California Institute of Technology, 4800 Oak Grove Drive, Pasadena, CA 91109, USA \\
$^{3}$School of Physics and Astronomy, Cardiff University, Queens Buildings, The Parade, Cardiff CF24 3AA, UK
}

\date{Accepted 2026 October 14. Received 2026 September 10; in original form 2026 May 03.}

\pubyear{\the\year{}}

\begin{document}
\label{firstpage}
\pagerange{\pageref{firstpage}-\pageref{lastpage}}
\maketitle

\begin{abstract}
The Habitable Worlds Observatory (HWO) will characterize Earth-like exoplanets from low-resolution reflected-light spectra. To assess how atmospheric evolution shapes what can be recovered, we trained a convolutional neural network on 109,945 spectra generated with the Planetary Spectrum Generator for Archean, Proterozoic, and Modern atmospheres under LUVOIR-B and HabEx/SS observing assumptions, separating the contributions of observational noise and model variability. We find that recovery depends jointly on abundance and spectral structure. Across eras, O$_2$ recovery weakens in Proterozoic atmospheres and CH$_4$ becomes reliable only when abundant, while an isolated UV band yields robust Modern O$_3$ recovery ($R^2\simeq0.98$) despite poor Modern CH$_4$ recovery ($R^2=0.32$-$0.35$). The wavelength analysis reinforces this interpretation. Without being supplied feature locations, the network assigns $58$-$71\%$ of total attribution to visible wavelengths, while the ultraviolet remains decisive for O$_3$ and the near-infrared for CH$_4$ and CO$_2$. Aggregate retrieval leverage declines beyond approximately $1.7\,\mu$m, supporting a practical long-wavelength boundary in this setup. These wavelength dependencies also shape detectability. At baseline noise, $83$-$85\%$ of Modern O$_2$ cases satisfy a $3\sigma$ criterion, while Proterozoic O$_2$ remains below $50\%$ when noise is halved and Modern CH$_4$ remains below $3\sigma$ as observational noise approaches zero because model uncertainty caps its significance. Together, these results imply molecule-specific priorities for HWO trade studies. Low-abundance O$_2$ retains considerable leverage from improved observational quality while motivating better inference, whereas at the 1,000-hour benchmark O$_3$ and Modern CH$_4$ show predominantly model-associated uncertainty in these networks, motivating broader training coverage and improved uncertainty calibration.
\end{abstract}




\section{Introduction}

Since the discovery of the first exoplanet orbiting a Sun-like star in 1995 \citep{mayorqueloz1995}, over 6{,}400 exoplanets have been confirmed\footnote{For the latest count, see the NASA Exoplanet Archive: \url{https://exoplanetarchive.ipac.caltech.edu/}.}. These planets span a wide range of sizes, orbital architectures, and bulk compositions, motivating observational and theoretical efforts to understand their formation and evolution \citep{Morbidelli2016}.

As the exoplanet census has expanded, atmospheric characterization has become a central objective \citep{Charbonneau2002, Charbonneau2005, Gibson2010, Brogi2012, deming2017, estrela2022}. For small, temperate planets, atmospheric signals are intrinsically faint and remain challenging to measure. Consequently, much of the current literature has focused on giant planets using transmission spectroscopy (e.g., \citealt{Kreidberg2014, Tsiaras2019, Robinson2017, JWST2023}), which probes the planetary limb during transit but requires near-edge-on geometries and is observationally biased toward short-period systems.

Direct imaging provides a complementary route by spatially separating planet and star, enabling reflected-light spectroscopy of the planetary disk. Rapid advances in high-contrast imaging, coronagraphy, and wavefront control indicate that the contrasts required to characterize nearby rocky planets are within reach of next-generation facilities \citep{Damiano2022, Damiano2023, Robinson2023, Currie2023}. A key upcoming mission in these efforts is NASA's \textit{Habitable Worlds Observatory} (HWO), conceived as a future flagship mission building on concepts from the Habitable Exoplanet Observatory (HabEx) and the Large UV/Optical/InfraRed Surveyor (LUVOIR) \citep{LUVOIR2019, Gaudi2020, Coston2024, Feinberg2026}. The HWO architecture is still under development, with different concepts being considered for aperture, pupil geometry, starlight suppression, and instrument configuration. In this context, LUVOIR-B and HabEx/SS provide two contrasting heritage architectures that span part of the design space relevant to HWO, without implying that either represents the final observatory design. Following recommendations from the Decadal Survey on Astronomy and Astrophysics 2020 (Astro2020) \citep{astro2020}, HWO is designed to detect and characterize on the order of a few dozen potentially habitable rocky planets, obtaining spectra from the ultraviolet (UV) through the visible (VIS) and into the near-infrared (NIR). This wavelength coverage includes diagnostically important absorption bands of O$_2$ and O$_3$ \citep{Damiano2023}.

In this context, Earth provides the most detailed reference for interpreting the atmospheric diversity expected among rocky exoplanets (e.g., \citealt{arney2016pale, Rugheimer2018, Kofman2024}). While present-day Earth is often summarized as an N$_2$-dominated atmosphere with O$_2$, trace CO$_2$, and variable H$_2$O \citep{kopparapu2013habitable, schwieterman2018exoplanet}, geological records show that Earth's atmospheric composition evolved substantially over time \citep{lyons2014rise}.  During the Archean eon (4.0-2.5 Ga), intense volcanic activity reshaped Earth's surface, while the atmosphere was dominated by N$_2$, CH$_4$, and CO$_2$ with negligible O$_2$ \citep{Catling2020}. The subsequent Proterozoic eon (2.5 Ga-500 Myr ago) saw the Great Oxidation Event (GOE), driven by photosynthetic microorganisms, which dramatically increased atmospheric O$_2$ levels \citep{Planavsky2014}. Although Proterozoic O$_2$ and O$_3$ levels were insufficient to produce strong VIS band absorption features, they significantly influenced the UV regime \citep{Damiano2023}. The Phanerozoic eon (last 500 Myr) includes the Cambrian explosion and a pivotal diversification of complex life \citep{Marshall2006, Catling2017}, culminating in the modern N$_2$-O$_2$-dominated atmosphere. These three epochs therefore serve as physically motivated templates for evolving Earth analogues targeted by future reflected-light surveys.

A persistent challenge for such surveys is to infer atmospheric and planetary parameters from low-resolution spectra in a high-dimensional parameter space \citep{Changeat2023}. In practice, spectrum interpretation typically combines forward radiative-transfer and atmospheric-chemistry modeling with inverse retrieval via Bayesian inference frameworks that compare forward models to data to estimate posterior distributions over the parameters (e.g., \citealt{Madhusudhan2009, Waldmann2015, Lavie2017, Damiano2020}). Although these approaches provide posterior distributions based on physically motivated forward models, their computational cost increases with model dimensionality and is incurred separately for each spectrum, making repeated retrievals across large synthetic populations and multiple observing scenarios impractical.

Motivated by these computational constraints, Machine Learning (ML) methods have been increasingly explored as scalable alternatives for exoplanet spectroscopy, both as fast surrogates and as inference models. In the context of atmospheric retrieval, early work by \citet{Waldmann2016} introduced RobERt, a deep-belief network for classifying exoplanet spectra, followed by ExoGAN \citep{Zingales2018} and supervised regressors with uncertainty estimates \citep{Forestano2025}. Convolutional Neural Network (CNN) and other deep architectures have also been applied directly to atmospheric-parameter inference from transmission spectra, including comparisons of different neural architectures and evaluations against nested-sampling retrievals \citep{Yip2021, ArdevolMartinez2022}. A complementary line of research focuses on learning flexible posterior densities. Normalizing flows have become prominent density estimators in simulation-based inference and related variational approaches, including amortized neural posterior estimation \citep{Vasist2023}, sequential neural posterior estimation \citep{ArdevolMartinez2024}, variational inference coupled to a differentiable forward model \citep{Yip2024}, and flow-matching posterior estimation combined with importance-sampling validation \citep{Gebhard2025}. In parallel, ML has been applied to complementary spectroscopy tasks, including analytical and data-driven dimensionality reduction for transit spectra \citep{Matchev2022, Matchev2022b} and anomaly detection to flag spectra with unusual chemical signatures \citep{Forestano2023}. Focusing specifically on rocky terrestrial planets, \citet{Zorzan_2025} presented a large ML-ready dataset of synthetic planetary systems generated with NASA's Planetary Spectrum Generator (PSG) within the INARA framework.

Building on this foundation and to support the science goals of the HWO, we present a 1D CNN trained on 109,945 synthetic reflected-light spectra of Earth-analogue planets spanning Archean, Proterozoic, and Modern atmospheric compositions. Spectra are generated with PSG under noise models consistent with LUVOIR-B and HabEx/SS design specifications, placing the retrieval problem in a mission-relevant context where wavelength coverage, spectral resolution, and instrumental noise jointly determine the accessible information. The network simultaneously retrieves six atmospheric mixing ratios (CH$_4$, CO$_2$, H$_2$O, N$_2$, O$_2$, O$_3$) and four planetary parameters (radius, gravity, surface temperature, and surface pressure). Relative to prior deep-learning approaches, the present framework incorporates three elements aimed specifically at mission relevance: mission-consistent noise from LUVOIR-B and HabEx/SS forward models; retrieval optimization for O$_2$ and O$_3$; and interpretability through integrated-gradient attribution, which identifies the spectral features most strongly associated with each prediction within the trained network. O$_2$ and O$_3$ are emphasized because they are widely studied potential oxygen-related biosignature gases with absorption features accessible to UV-VIS reflected-light spectroscopy \citep{DesMarais2002, schwieterman2018exoplanet}. However, detecting either species would not by itself establish biological activity, because both can be produced or accumulated through abiotic pathways and must therefore be interpreted in their planetary and stellar context \citep{Meadows2018}. By explicitly linking retrieval performance to bandpass, resolution, and noise assumptions, these results are intended not only as a retrieval demonstration but also as quantitative input for HWO trade studies, target-yield planning, and the broader question of where remaining limitations are set by noise, spectral coverage, or model degeneracy.

The paper is structured as follows. Section \ref{sec:2} details the methodology for generating synthetic reflected-light spectra and planetary configurations. Section \ref{sec:3} describes the CNN architecture, training and evaluation. Section \ref{sec:4} presents the retrieval results and discussion. Section \ref{sec:5} addresses the limitations of the framework, and Section \ref{sec:conclusion} presents the conclusions and their implications.

\section{Training Data Generation} \label{sec:2}

The synthetic reflection spectra dataset was generated using NASA’s PSG\footnote{\url{https://psg.gsfc.nasa.gov}} \citep{psg2018}, a radiative transfer model that synthesizes planetary spectra by coupling atmospheric radiative transfer equations with high-resolution spectroscopic databases. For efficient large-scale data generation, we employed a locally hosted PSG instance running within a Docker container, following the implementation guidelines provided by \citet{villanueva2022fundamentals}.

Each spectrum is defined by a configuration dictionary that specifies stellar, orbital, atmospheric, surface, and instrument parameters\footnote{A full parameter reference is available at \url{https://psg.gsfc.nasa.gov/helpapi.php\#parameters}.}. For every configuration, PSG returns the reflected-light spectrum and an instrument-noise estimate (and, optionally, intermediate radiative-transfer products such as transmittance profiles). Building the training set involves two integrated steps: (1) physical screening, in which we apply first-order criteria for atmospheric retention and the thermodynamic stability of surface liquid water; and (2) sample realization, in which an epoch-dependent atmospheric template is converted to a pressure-weighted isoabundance composition, perturbed in log-ratio space, and combined with sampled stellar, orbital, planetary, and observational parameters for the LUVOIR-B and HabEx/SS exposure scenarios. The subsections that follow describe each component in detail.

\subsection{Habitable zone for F- and G-type stars}

A key driver in the search for exoplanets is the discovery of potentially habitable worlds. However, the concept of \textit{habitability} is complex and influenced by several factors, including the characteristics of the host star, system dynamics, and the planet's specific composition \citep{kane2021planetary}. Our understanding of habitability is primarily based on conditions that support life on Earth: the presence of liquid water as a solvent, a suitable temperature range, an energy source, and essential biogenic elements \citep{owen1980search, cockell2016habitability}.

The habitable zone (HZ), the region around a star where rocky planets can sustain surface liquid water, provides a first-order constraint for identifying potentially habitable worlds \citep{kasting1993habitable, brack1993liquid, kaltenegger2017}. \citet{kopparapu2014habitable} define both conservative and optimistic HZ boundaries for planets around stars with $2600\,\mathrm{K}\leq T_{\rm eff}\leq7200\,\mathrm{K}$. The conservative boundaries are associated with the runaway-greenhouse and maximum-greenhouse limits, whereas the optimistic HZ is bounded by the Recent Venus and Early Mars empirical limits. In this work, we adopt the optimistic HZ, using Recent Venus ($S_{\rm RV}$) as the inner boundary and Early Mars ($S_{\rm EM}$) as the outer boundary. Both limits are expressed as fourth-order polynomials in $T_0\equiv T_{\rm eff}-5780\,\mathrm{K}$,
\begin{align}
S_{\text{EM}} &= 0.32 + (5.547 \times 10^{-5} )\,T_0 + (1.526 \times 10^{-9})\,T_0^2 \nonumber \\
&\qquad - (2.874 \times 10^{-12} )\,T_0^3 - (5.011 \times 10^{-16} )\,T_0^4, \\
S_{\text{RV}} &= 1.776 + (2.136 \times 10^{-4})\,T_0 + (2.533 \times 10^{-8})\,T_0^2 \nonumber \\
&\qquad - (1.332 \times 10^{-11} )\,T_0^3 - (3.097 \times 10^{-15} )\,T_0^4.
\end{align}

To compute orbital distances within this zone, we select F- and G-type dwarfs with equal probability and sample $T_{\rm eff}$ uniformly over $5990$-$7200\,$K for F stars and $5380$-$5930\,$K for G stars. Instead of sampling the stellar radius independently, we conditionally interpolate $\log R_\star$ as a function of $T_{\rm eff}$ along the Pecaut-Mamajek mean dwarf sequence \citep{mamajek2019modern}. These spectral types are favored because their long main-sequence lifetimes permit biological evolution, unlike the short-lived O-, B-, and A-type stars \citep{cuntz2016exobiology}. A preliminary calculation that also included K-type hosts produced no K-star spectra satisfying the adopted signal-to-noise ratio (SNR) requirement; the production sampling was therefore restricted to F and G stars. Stellar luminosity is then obtained via the Stefan-Boltzmann relation, and the HZ inner and outer boundaries follow as,
\begin{equation}
d = \left( \frac{L_\star/L_\odot}{S} \right)^{1/2},
\end{equation}
where $S$ takes $S_{\rm RV}$ for the inner edge and $S_{\rm EM}$ for the outer edge. The resulting distance interval defines the orbital-radius prior for our synthetic planet catalogue, from which the semi-major axis is drawn uniformly in distance.

\subsection{Criteria for sustaining atmosphere and habitability}\label{sec:2.2}

Beyond the geometric limits of the HZ, a rocky planet must satisfy additional conditions for liquid water to persist on its surface. Foremost among them is the presence of an atmosphere, which regulates the planet's radiative‐convective energy budget by coupling stellar insolation to thermal emission and interior heat flow \citep{seager2013,keles2018}. 

We therefore consider three critical factors governing atmospheric retention under habitable conditions: (1) surface pressure ($P_{\rm surf}$), estimated via hydrostatic scaling combined with an empirical mass-radius (M-R) relation; (2) surface temperature ($T_{\rm surf}$), governed primarily by incident stellar flux and atmospheric opacity; and (3) the cosmic shoreline \citep{zahnle2017}, an empirical boundary that links stellar irradiation and planetary gravity to the long‐term retention or loss of volatile envelopes. The remainder of this subsection quantifies each of these parameters and defines the acceptance criteria applied.

\subsubsection{Surface pressure}

Direct measurements of exoplanetary surface pressure are not yet feasible, so we rely on first‑order hydrostatic scaling coupled to empirical M-R relations. Observationally derived M-R trends are usually expressed as piecewise power laws that track compositional regimes \citep[e.g.,][]{hatzes2015, otegi2020}. For terrestrial bodies, \citet{chen2017} find,
\begin{equation}\label{4}
R_{\rm p} \propto M_{\rm p}^{\,0.279 \pm 0.009},
\end{equation}
a result corroborated by the more recent analysis of \citet{muller2024}, with differences confined to the third decimal place for $M_{\rm p} \lesssim 4.4\,M_\oplus$.

For each planetary proposal, we draw a single mass-radius exponent $q\sim\mathcal{U}(0.270,0.288)$ and calculate
\begin{equation}\label{eq:mass_radius}
\frac{M_{\rm p}}{M_\oplus}=\left(\frac{R_{\rm p}}{R_\oplus}\right)^{1/q}.
\end{equation}
Following the hydrostatic scaling of \citet{mcintyre2023}, the surface pressure in Earth units obeys
\begin{equation}
\frac{P_{\rm surf}}{P_\oplus} = \left(\frac{M_{\rm p}}{M_\oplus}\right)^{2} \left(\frac{R_\oplus}{R_{\rm p}}\right)^{4}.
\end{equation}
Substituting equation \ref{eq:mass_radius} into the previous expression yields
\begin{equation}\label{6}
\frac{P_{\rm surf}}{P_\oplus} = \left(\frac{R_{\rm p}}{R_\oplus}\right)^{2/q-4},
\end{equation}
where $P_\oplus=1013.25\,$mbar. A single draw of $q$ propagates to the planetary mass, gravity, escape velocity, and pressure exponent, so that all four remain mutually consistent for a given planet. For each proposed realization, we sample the planetary radius from $R_{\rm p}/R_\oplus\sim\mathcal{U}(0.3,1.23)$.
The bounds follow the rocky-planet interval adopted by \citet{mcintyre2023}, spanning the range from the smallest confirmed rocky exoplanet around a main-sequence star \citep{Barclay2013} up to the empirical transition between terrestrial and volatile-rich worlds \citep{chen2017}. Within this interval equation \ref{6} yields surface pressures from $\sim 0.02$ to $\sim2$ bar.

\subsubsection{Surface temperature}

Surface temperature is also a key parameter for establishing the phase stability of liquid water. Although $T_{\rm surf}$ can in principle be obtained directly from sophisticated radiative‑convective or general‑circulation models \citep{way2017,phillips2020}, such treatments are computationally prohibitive for large-scale sampling. Instead, we adopt a semi‑analytical approach that rescales the planet's equilibrium temperature ($T_{\rm eq}$) by a simple greenhouse‑effect factor.

The equilibrium temperature represents the blackbody temperature of a planet in radiative balance with its host star \citep{seager2010}:
\begin{equation}\label{eq:Teq}
T_{\rm eq} = \left[\frac{(1-A_B)\,(I_\star/I_\oplus)\,I_\oplus}{4\sigma}\right]^{1/4},
\end{equation}
where $A_B$ is the Bond albedo, $\sigma$ is the Stefan-Boltzmann constant, $I_\oplus \simeq 1361\,\mathrm{W\,m^{-2}}$ is the solar irradiance at Earth, and the stellar insolation relative to Earth is,
\begin{equation}\label{eq:insolation}
\frac{I_\star}{I_\oplus} = \frac{L_\star}{L_\odot} \left(\frac{a_\oplus}{a_{\rm p}}\right)^{2},
\end{equation}
with $a_{\rm p}$ the planet's semi-major axis. To convert $T_{\rm eq}$ to a surface value, we follow the idealized greenhouse prescription of \citet{seager2010},
\begin{equation} \label{eq:Tsurf}
T_{\rm surf} = \left(\frac{2}{2-\varepsilon_{\rm IR}}\right)^{1/4} T_{\rm eq},
\end{equation}
where $\varepsilon_{\rm IR}$ is an effective atmospheric infrared emissivity parameter. We sample $A_B\sim\mathcal{U}(0.06,0.96)$ and $\varepsilon_{\rm IR}\sim\mathcal{U}(0,1)$ independently: $A_B$ controls the absorbed shortwave radiation, whereas $\varepsilon_{\rm IR}$ parameterizes longwave trapping. This reproduces two limiting cases: (i) no atmospheric greenhouse effect ($\varepsilon_{\rm IR} \to 0$), for which $T_{\rm surf} = T_{\rm eq}$; and (ii) a fully opaque greenhouse ($\varepsilon_{\rm IR} \to 1$), for which $T_{\rm surf}=2^{1/4}T_{\rm eq}$. For Earth-like conditions ($A_B\simeq0.3$, $\varepsilon_{\rm IR} \approx 0.78$, $I_\star/I_\oplus=1$), equation \ref{eq:Tsurf} returns $T_{\rm surf}\simeq288\,$K, consistent with observations \citep{delgenio2019,mcintyre2023}.

The albedo interval captures the diversity of planetary reflectivities; note that low-albedo planets are intrinsically harder to detect because they reflect less starlight \citep{Gaudi2020}.

After drawing $A_B$ and $\varepsilon_{\rm IR}$ independently, we check whether the resulting $T_{\rm surf}$ lies between the water freezing and boiling points at the prevailing surface pressure. The boiling curve is obtained from the Clausius-Clapeyron relation, while the freezing point is fixed at 273.15\,K whenever $P_{\rm surf}>P_{\rm triple}=611.657\,$Pa. For $P_{\rm surf}\le P_{\rm triple}$, liquid water cannot exist as a thermodynamically stable phase, and such configurations are rejected.

For every atmosphere that passes the temperature filter, we retain the native 60-layer P-T structure of its geological template and rescale it to the sampled surface boundary conditions. If $P_j^{(0)}$ and $T_j^{(0)}$ are the template values and $P_{\rm surf}^{(0)}$ and $T_{\rm surf}^{(0)}$ are the corresponding surface values, then
\begin{align}
P_j &= P_j^{(0)}\frac{P_{\rm surf}}{P_{\rm surf}^{(0)}},
\label{eq:pressure_rescale}\\
T_j &= T_j^{(0)}+T_{\rm surf}-T_{\rm surf}^{(0)}.
\label{eq:temperature_rescale}
\end{align}
This transformation preserves the relative pressure spacing and the vertical temperature structure of the template while imposing the new surface pressure and temperature.

\subsubsection{Cosmic Shoreline}

We finalize the habitability filter by evaluating whether a planet's atmosphere can withstand stellar‑driven escape. An empirical threshold known as the cosmic shoreline represents this trade‑off between incoming stellar irradiation and a planet's gravitational binding energy \citep{zahnle1998,zahnle2013,zahnle2017}. In its most basic formulation, this shoreline is described by a power-law relation:
\begin{equation}\label{eq:shoreline}
I_\star = k\,v_{\rm esc}^{\,4},
\end{equation}
where $I_\star$ is the dimensionless insolation relative to Earth (equation~\ref{eq:insolation}), $v_{\rm esc}=\sqrt{2GM_{\rm p}/R_{\rm p}}$ is the escape velocity, and $k=5\times10^{-16}\,({\rm m\,s^{-1}})^{-4}$ is the empirical normalization adopted from \citet{mcintyre2023}. Planets with $I_\star<k\,v_{\rm esc}^{\,4}$ are expected to be more favorable for long-term atmospheric retention, whereas planets above this boundary are more susceptible to atmospheric loss.

For each simulated planet, we compute $I_\star$ from its orbital distance and stellar luminosity (equation \ref{eq:insolation}) and compare the value to the threshold given by equation \ref{eq:shoreline}. Approximately two‑thirds ($\simeq66\%$) of our configurations satisfy the retention criterion.

\subsection{Geological epochs}

For each geological epoch, we adopt literature-based vertically resolved profiles of composition, pressure, and temperature as templates. For every accepted planetary realization, the P-T profile is rescaled to match the sampled surface boundary conditions $(P_{\rm surf},T_{\rm surf})$: pressure is scaled multiplicatively and temperature is shifted additively, so that the pressure ratios and temperature differences between layers are preserved exactly. The vertical composition profiles are treated differently. They are not passed to the forward model layer by layer, but are reduced to a single pressure-weighted baseline abundance per species (Section~\ref{sec:mixing_ratio}). The resulting abundances are then perturbed and applied uniformly to all atmospheric layers.

The Archean atmospheric profile is based on the photochemical model from \citet{arney2016pale}. A defining feature of the Archean atmosphere is the organic haze formed via methane photolysis, which polymerizes into fractal hydrocarbon aerosols. We adopt the baseline profiles corresponding to CH$_4$/CO$_2 = 0.21$; however, we do not include explicit haze or aerosol optical properties in PSG, using the gaseous composition of the profile as the baseline for the pressure-weighted isoabundance treatment described in Section~\ref{sec:mixing_ratio}.

The Proterozoic atmospheric data were extracted from \citet{kawashima2019}, specifically for the post-GOE period ($\sim$2.33 Ga). During this epoch, O$_2$ was approximately $10^{-2}$ times the present atmospheric level (PAL), corresponding to a volume mixing ratio of $\sim 2 \times 10^{-3}$. Methane concentrations, though poorly constrained by geological records, are hypothesized to have been significantly higher than on the modern Earth.

Modern atmospheric profiles are obtained from the NASA/MERRA-2 reanalysis dataset \citep{gelaro2017merra2}, which assimilates satellite and ground-station observations into 72 vertical layers at a horizontal resolution of approximately $0.5^\circ$. PSG refines these profiles to 1 km vertical resolution using embedded topographic information and hydrostatic equilibrium assumptions. For atmospheric species not included in MERRA-2, PSG employs standard abundance profiles to fill data gaps; further details are provided in \citet{gelaro2017merra2} and \citet{villanueva2022fundamentals}.

\subsubsection{Mixing ratio} \label{sec:mixing_ratio}

To capture compositional degeneracies that often affect atmospheric retrievals, we perturb the baseline gas abundances associated with each epoch template. Before applying the perturbations, we replace the vertically varying abundance profile of each species $i$ by its pressure-weighted mean,
\begin{equation}\label{eq:pressure_weighted_abundance}
\overline{Y}_i=
\frac{\displaystyle\int_{P_{\rm top}}^{P_{\rm surf}}Y_i(P)\,dP}
{\displaystyle P_{\rm surf}-P_{\rm top}},
\end{equation}
evaluated by trapezoidal integration over the native pressure levels. Under hydrostatic equilibrium $\mathrm{d}P$ is proportional to the mass of the layer, so each level is weighted by the amount of gas it contains. This matters for vertically stratified species such as O$_3$, whose mixing ratio peaks in the stratosphere where the density is low. The pressure-weighted abundances of the non-N$_2$ gases are retained, and N$_2$ initially fills the remainder required for the abundances to sum to unity.

For each positive species $i$ other than N$_2$, we then perturb its abundance ratio relative to N$_2$ symmetrically in natural-log space. Defining $r_i^{(0)}=\overline{Y}_i/\overline{Y}_{\rm N_2}$, we draw
\begin{equation}\label{eq:12}
\ln r_i=\ln r_i^{(0)}+\delta_i,
\qquad \delta_i\sim\mathcal{U}(-2.5,2.5).
\end{equation}
This permits both enrichment and depletion relative to the template, with gas-to-N$_2$ ratio multipliers between $e^{-2.5}\simeq0.082$ and $e^{2.5}\simeq12.18$. The perturbed ratios are transformed into a closed composition using
\begin{equation}\label{eq:composition_closure}
Y_{\rm N_2}=\frac{1}{1+\sum_i r_i},
\qquad Y_i=r_iY_{\rm N_2},
\end{equation}
which guarantees non-negative abundances and $\sum_iY_i=1$. Species that are absent from an epoch template remain exactly zero. To keep the sampled compositions within the adopted geological family, we require $Y_{\rm N_2}\geq0.5\,\overline{Y}_{\rm N_2}$ and redraw proposals that violate this condition.

For the forward-model calculations, we adopt \emph{isoabundance} columns, i.e., the same accepted composition is applied uniformly across all atmospheric layers, while the pressure and temperature retain the rescaled epoch-dependent vertical structure. The mean molecular weight is recomputed from the normalized final composition. This separation disconnects composition from the P–T structure, so that correlations can be inferred from the training data instead of being hard-wired by design.

All positive non-N$_2$ gases present in each PSG template are perturbed, including gases that are not retrieval targets. In particular, CO and N$_2$O remain trace constituents within the controlled perturbations around the adopted geological templates and are not predicted by the network. The training population therefore contains no CO-rich or CO-dominated atmospheres. Because we do not impose equilibrium-chemistry constraints, this stochastic parameterization remains agnostic to specific reaction pathways and provides a controlled exploration of atmospheric variability around each geological template.

\subsection{Telescope properties} \label{sec:telescope_properties}

\begin{figure*}
\centering
\includegraphics[width=\textwidth]{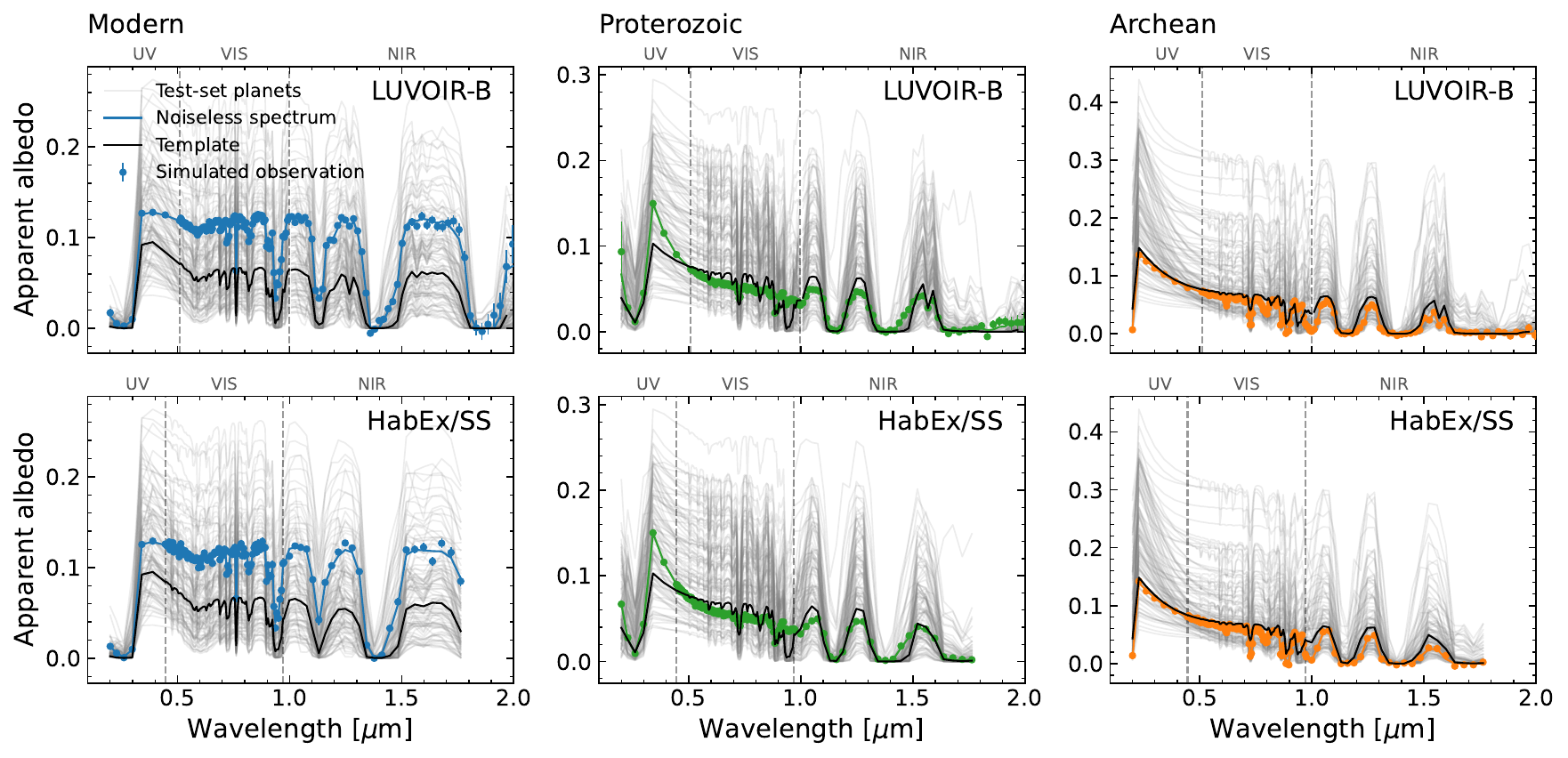}
\caption{Examples of test-set reflection spectra for Earth-like exoplanets in the Modern (left), Proterozoic (center), and Archean (right) eras, observed with LUVOIR-B (top row) and HabEx/SS (bottom row). Grey lines show 100 spectra drawn at random from the test set for each era; one of them is highlighted in the colour of its era. The thick colored line represents the noiseless apparent albedo $A(\lambda)$, while points with error bars correspond to $A_{\text{noisy}}(\lambda)$ as defined by equation \ref{eq:noisy_spectrum}. The thin black line is the unperturbed template atmosphere of each era, and dashed vertical lines mark the instrument channel boundaries. The same planets are shown in both rows, so panels within a column differ only in the instrument.}
\label{fig:1}
\end{figure*}

We generate synthetic observations for two mission concepts implemented in PSG: the 8 m coronagraphic LUVOIR‑B and the 4 m starshade‑assisted HabEx/SS. These configurations are not assumed to represent the final HWO architecture. Instead, they capture two contrasting points in the design heritage informing HWO, with different apertures and starlight-suppression strategies. Comparing them allows us to examine architecture-dependent retrieval behavior under internally consistent instrument models without forecasting the performance of a finalized HWO design. Both designs target a planet–star contrast of $10^{-10}$ and we adopt the default instrument parameters recommended in their respective mission reports \citep{LUVOIR2019,Gaudi2020}.

Regarding spectral coverage, LUVOIR-B observes in three bands: UV (0.20-0.515 $\mu$m, $R=7$), VIS (0.515-1.00 $\mu$m, $R=140$), and NIR (1.01-2.00 $\mu$m, $R=70$). HabEx/SS covers slightly shifted ranges: UV (0.20-0.45 $\mu$m, $R=7$), VIS (0.45-0.975 $\mu$m, $R=140$), and NIR (0.975-1.80 $\mu$m, $R=40$). PSG samples each spectrum using a boxcar kernel matched to the resolving power \citep{villanueva2022fundamentals} and applies internal oversampling to preserve spectral fidelity \citep{psg2018}.

A key difference between the two designs lies in starlight suppression. LUVOIR‑B's internal coronagraph imposes an inner working angle of $3.5\,\lambda/D$ \citep{checlair2021}, corresponding to $\sim45$ mas at $0.5\,\mu$m for $D=8$ m and limiting detection of planets at small angular separations. HabEx/SS avoids this wavelength-dependent constraint by using an external starshade, for which the effective inner working angle is approximately constant within each band (39, 58, and 104 mas in the UV, VIS, and NIR bands, respectively) \citep{Gaudi2020}.

Noise modeling in PSG includes throughput losses, zodiacal light, stellar leakage, and detector noise. For LUVOIR-B, we assume back‑illuminated CCDs with per‑pixel read noise of 0 e$^{-}$ (UV/VIS) and 2.5 e$^{-}$ (NIR), and dark noise of $3\times10^{-5}\,\text{e}^{-}\,\text{s}^{-1}\,\text{px}^{-1}$ (UV/VIS) and 0.002 e$^{-}\,\text{s}^{-1}\,\text{px}^{-1}$ (NIR). For HabEx/SS, which assumes photon-counting detectors in the UV/VIS channel, the corresponding values are 0.008 e$^{-}$ (UV/VIS) and 0.32 e$^{-}$ (NIR), with dark noise of $3\times10^{-5}\,\text{e}^{-}\,\text{s}^{-1}\,\text{px}^{-1}$ (UV/VIS) and 0.005 e$^{-}\,\text{s}^{-1}\,\text{px}^{-1}$ (NIR) \citep{checlair2021}. Readers can find detailed equations for each noise component in \citet{villanueva2022fundamentals}.

Instrument configuration flags in PSG enable independent activation of molecular, continuum, and stellar-absorption features. We allow all three for LUVOIR-B, whereas for HabEx/SS we disable the stellar absorption lines imprinted on the starlight reflected by the planet. Each simulation assumes a total exposure time of 1,000 h per planet, implemented as 3,600 integrations of 1,000 s. Following \citet{checlair2021}, we adopt this as an upper-limit deep-characterization benchmark for LUVOIR- and HabEx-class observatories, without assuming that the same allocation would apply to every target. Finally, planetary geometries are randomized across distances of 5-16 pc. Orbital orientations are sampled isotropically subject to $i\leq50^\circ$, by drawing $\cos i$ uniformly over $[\cos50^\circ,1]$, while orbital phase is sampled uniformly over $0$-$360^\circ$. We adopt zero obliquity, resulting in phase angles of approximately $40^\circ$-$140^\circ$. These geometrical parameters are used by PSG when generating each spectrum, but neither inclination, orbital phase, nor phase angle is supplied to the CNN during training or inference. The reported retrieval performance therefore averages over the sampled viewing geometries. Each planet realization is simulated at a single phase; joint retrievals from repeated observations of the same planet at multiple phases are not considered here.

\subsection{Sample selection} \label{sec:sample_selection}

The physical and observational filters described above greatly restrict the habitable parameter space. Approximately 66\% of the proposals fall below the cosmic shoreline, but only 14\% of those also satisfy the P-T requirements for surface liquid water, so that $\sim9\%$ of all proposals are accepted. In the remaining cases, configurations fall outside the liquid-water stability regime primarily because they yield surface temperatures too low to prevent freezing, even when atmospheric retention is permitted by equation \ref{eq:shoreline}. To construct a viable sample efficiently despite this low acceptance rate, we use a recursive sampling procedure. For each fixed stellar and atmospheric-composition draw, the sampling algorithm performs up to 200 iterations, drawing a new semi-major axis, planetary radius and associated mass-radius exponent, Bond albedo, and atmospheric infrared emissivity at each step. For each accepted draw, one that satisfies both the cosmic shoreline criterion and the liquid-water stability conditions, a PSG configuration file is generated and executed, returning the apparent-albedo spectrum $A(\lambda)$, the wavelength-dependent noise array $\sigma(\lambda)$, and the individual radiative sub-components.

Once a realization passes the physical habitability criteria, we then evaluate its observational feasibility. We impose a SNR quality cut before including a spectrum in our final library. For each spectral band (UV, VIS, NIR), we compute a percentile-based SNR designed to measure the contrast of spectral features against the noise floor:
\begin{equation}\label{eq:snr_pctl}
{\rm SNR}_{\rm pctl} \equiv \frac{S_{90}-S_{10}}{\sqrt{2}\,\sigma_{\rm rms}},
\end{equation}
where $S_{90}$ and $S_{10}$ are the 90th and 10th percentiles of $A(\lambda)$ computed over that band, and $\sigma_{\rm rms}$ is the root-mean-square of $\sigma(\lambda)$ over the same band. The $\sqrt{2}$ factor is a heuristic normalization that approximately accounts for the variance of a two-sample contrast under independent Gaussian noise. This percentile-based definition is less sensitive to rare spikes than a global max-min amplitude, which is especially relevant in the UV where occasional outliers can inflate the estimated signal. We retain only spectra for which $\text{SNR}_{\rm pctl} > 3$ in all bands, ensuring that the spectral amplitude is well above the noise floor. For every sample that passes this cut, the noisy spectrum used for training, $A_{\rm noisy}(\lambda)$, is produced by drawing from a normal distribution centered on the noiseless apparent albedo with the standard deviation at that wavelength,
\begin{equation}\label{eq:noisy_spectrum}
A_{\text{noisy}}(\lambda) \sim \mathcal{N}(A(\lambda), \sigma(\lambda)).
\end{equation}

After applying all requirements, the final dataset comprises 109,945 synthetic spectra distributed as follows: 33,518 (30.5\%) for the Archean era, 36,996 (33.6\%) for the Proterozoic era, and 39,431 (35.9\%) for the Modern era. All retained cases correspond to planets orbiting F- and G-type stars, which account for 81.1\% and 18.9\% of the final sample, respectively. Examples of these synthetic spectra are shown in Fig. \ref{fig:1}.

\begin{figure}
    \centering
    \includegraphics[width=\columnwidth]{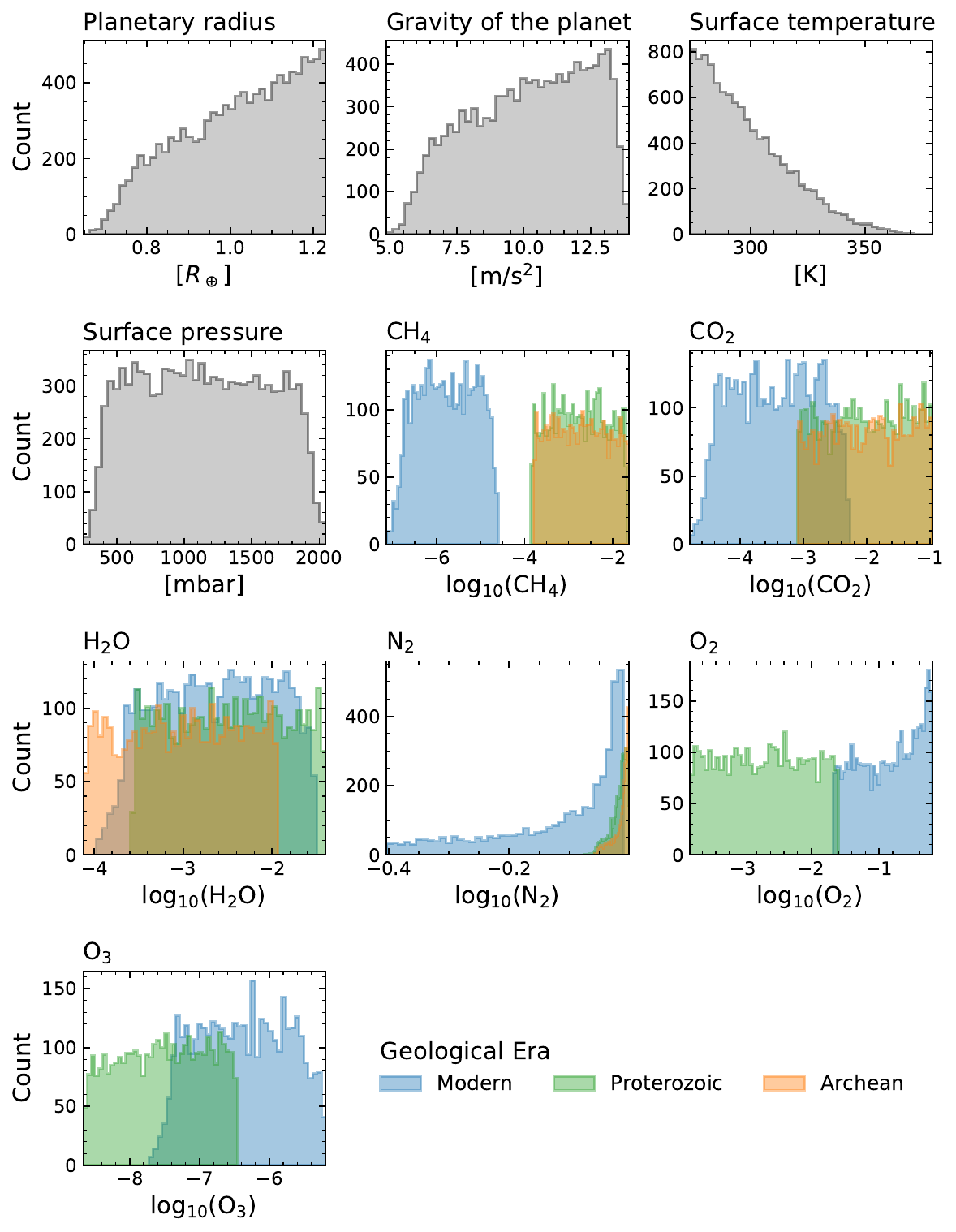}
    \caption{Distributions of planetary parameters and atmospheric mixing ratios in the final dataset. Grey histograms show the post-selection distributions of the physical parameters. Colored histograms show the epoch-dependent distributions of gas mixing ratios (in $\log_{10}$ space) where blue, green, and orange correspond to Modern, Proterozoic, and Archean samples, respectively.}
    \label{fig:distribution}
\end{figure}

Although planetary parameters were initially sampled from uniform priors according to the equations in previous subsections, the filters for habitability, atmospheric retention, and data quality reshaped these into the non-uniform distributions shown by the gray histograms in Fig. \ref{fig:distribution}. Planetary radius was initially sampled uniformly over $[0.3,1.23]R_\oplus$. No separate cutoff at $0.62R_\oplus$ was imposed; this is the smallest radius surviving the joint selection criteria in the regenerated data set. The final ranges for these physical parameters are listed in Table \ref{tab:1}.

\begin{table}
\centering
\caption{Final ranges of planetary physical parameters after sample selection.}
\label{tab:1}
\begin{tabular}{lcc}
\hline
{Parameter}              & {Min} & {Max} \\ \hline
Planetary radius ($R_{\oplus}$) & 0.62 & 1.23 \\
Planet gravity (m/s$^{2}$)      & 4.72 & 13.94 \\
Surface temperature (K)         & 273.2 & 381.4 \\
Surface pressure (mbar)         & 235 & 2049 \\ \hline
\end{tabular}
\end{table}

The atmospheric compositions were generated using the pressure-weighted, symmetric log-ratio perturbation scheme described in equation \ref{eq:12}, centered on epoch-dependent baseline abundance ratios relative to N$_2$. The resulting mixing-ratio distributions for each geological era are displayed by the colored histograms in Fig. \ref{fig:distribution}, while their minimum and maximum values are reported in Table \ref{tab:2} (with O$_2$ and O$_3$ absent in the Archean era).

\begin{table}
\centering
\caption{Final mixing-ratio ranges by geological epoch after sample selection.}
\label{tab:2}
\begin{tabular}{lcccccc}
\hline
 & \multicolumn{2}{c}{{Archean}} & \multicolumn{2}{c}{{Proterozoic}} & \multicolumn{2}{c}{{Modern}} \\
\cline{2-7}
{Parameter} & {Min} & {Max} & {Min} & {Max} & {Min} & {Max} \\
\hline
$\log_{10}$(CH$_4$) & $-3.84$ & $-1.63$ & $-3.89$ & $-1.67$ & $-7.14$ & $-4.60$ \\
$\log_{10}$(CO$_2$) & $-3.09$ & $-0.96$ & $-3.10$ & $-0.96$ & $-4.79$ & $-2.25$ \\
$\log_{10}$(H$_2$O) & $-4.14$ & $-1.94$ & $-3.59$ & $-1.39$ & $-4.03$ & $-1.49$ \\
$\log_{10}$(N$_2$)  & $-0.06$ & $-0.00$ & $-0.07$ & $-0.00$ & $-0.40$ & $-0.01$ \\
$\log_{10}$(O$_2$)  & \multicolumn{2}{c}{Absent} & $-3.81$ & $-1.60$ & $-1.68$ & $-0.22$ \\
$\log_{10}$(O$_3$)  & \multicolumn{2}{c}{Absent} & $-8.68$ & $-6.46$ & $-7.74$ & $-5.19$ \\
\hline
\end{tabular}
\end{table}

\section{Neural Network Training} \label{sec:3}

Our models were implemented, trained, and tested using the TensorFlow framework \citep{tensorflow2015}. In this section, we outline the motivation behind selecting a 1D CNN architecture and describe the data organization strategy used to maximize training efficiency.

\subsection{1D CNN}

Fully connected networks lack an inductive bias for local structure and therefore do not naturally exploit correlations between adjacent inputs. Reflection spectra, however, are inherently structured, that is, neighboring wavelengths are correlated, and this local continuity shapes patterns such as absorption bands and scattering slopes. To exploit this structure, we use 1D CNNs \citep{Rumelhart1985,LeCun1989,Kiranyaz2021}. A convolutional kernel learns to respond to specific spectral patterns, and because the same kernel is applied across the wavelength axis, those patterns can be detected even under small spectral shifts introduced by physical variability, a property known as \textit{translation-equivariance}. Strided convolutions then aggregate neighboring activations, enabling the model to emphasize the general shape of a feature instead of its precise position in each individual bin. This representation is particularly useful when atmospheric signatures are broad, partially overlapping, or modulated by continuum effects.

\begin{figure*}
\centering
\includegraphics[width=0.9\textwidth]{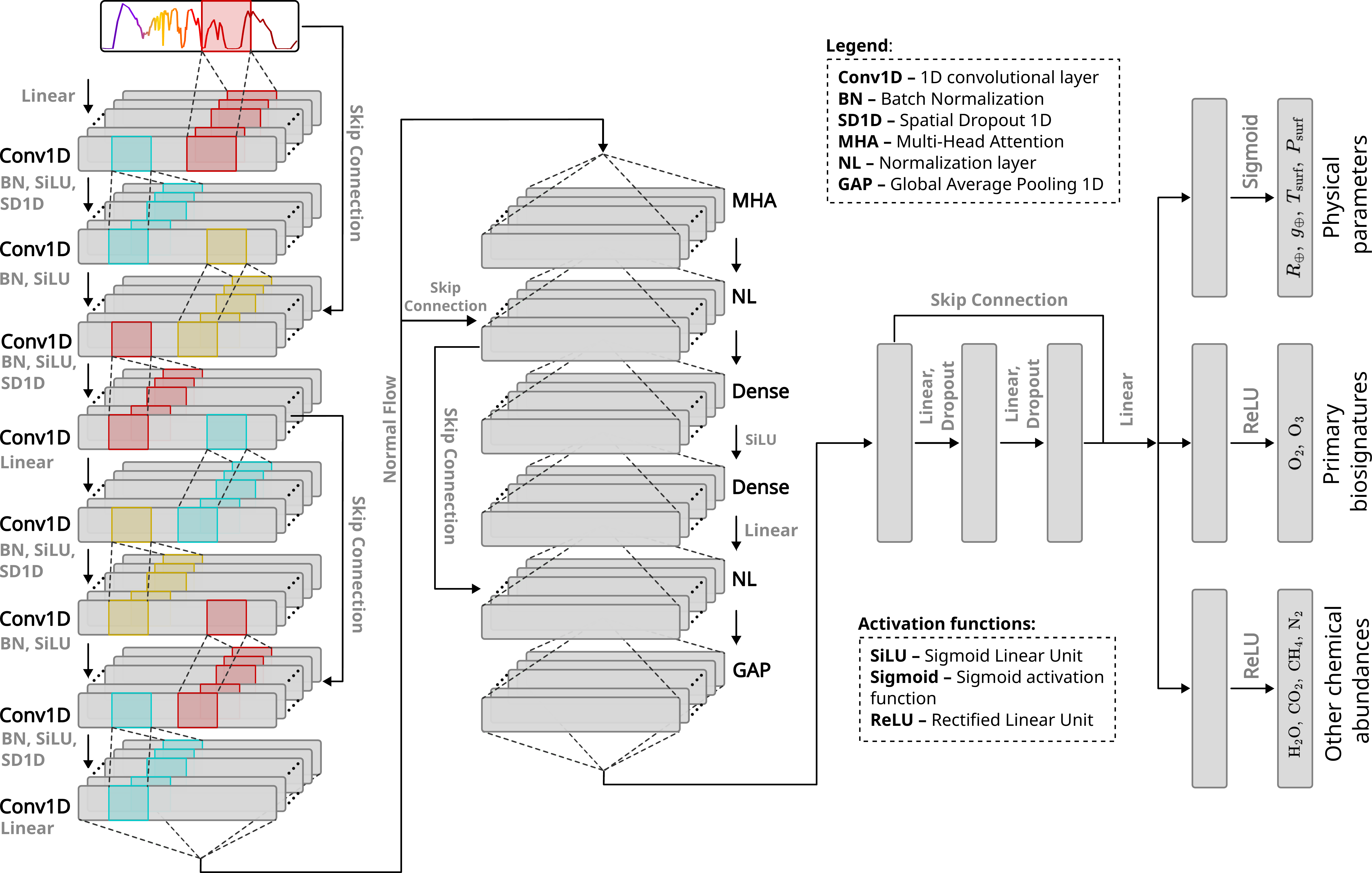}
\caption{Architecture of 1D CNN retrieval model. The input spectrum passes through residual convolutional blocks and a multi-head self-attention layer before being compressed by global average pooling into a latent vector. Three fully connected layers then branch into task-specific output heads for physical parameters and chemical abundances.} 
\label{fig:architecture}
\end{figure*}

The choice of a 1D CNN is also methodological. Traditional Bayesian retrieval remains the gold standard for detailed posterior estimation, but it is computationally expensive when the goal is to analyze large synthetic populations, perform repeated noise experiments, or support mission-level trade studies across many observing assumptions. At the other extreme, more flexible architectures such as Foundation models \citep{Bommasani2021} are powerful but computationally heavier than necessary for short, contiguous spectral inputs whose dominant information content is largely local. For the present application, a 1D CNN is well suited to the ordered wavelength structure of the spectra, allowing the model to learn both localized absorption features and broader continuum variations while retaining efficient inference. The trained models contain $9.2\times10^{4}$ parameters for LUVOIR-B and $4.9\times10^{5}$ for HabEx/SS. On a single CPU core, they predict all ten parameters in approximately $170$ and $428\,\mu$s per spectrum, respectively, allowing their corresponding test sets to be evaluated in a few seconds. We use this computational efficiency to perform controlled, large-scale comparisons across geological epochs and observing configurations.

We train the 1D CNN as a multi-output regression problem. At each iteration, the network parameters are updated by minimizing the mean absolute error (MAE) between the predicted and true target values. We use MAE instead of mean squared error because it is less affected by a few very large residuals, making it a more robust loss function for a dataset that covers a wide range of atmospheric conditions.

\subsection{Network architecture and training}

Fig. \ref{fig:architecture} summarizes the CNN architecture adopted in this work, whose dataset structure is described in Appendix \ref{app:dataset_structure} and \ref{app:A}. The same layout is used for both telescope concepts; only the input length differs due to the instrument-dependent wavelength sampling. The HabEx/SS input is a single concatenated sequence of length $L_{\mathrm{H}}=141$ (7 UV, 109 VIS, 25 NIR bins), while the LUVOIR-B input has length $L_{\mathrm{L}}=151$ (8 UV, 94 VIS, 49 NIR bins).

The input sequence is first processed by two residual convolutional blocks \citep{He2016}. In each block, a shortcut path is combined by addition with the transformed convolutional branch, allowing the block to learn a residual correction to its input instead of constructing an entirely new mapping. This construction facilitates optimization and gradient propagation through the network. The convolutional layers primarily extract patterns from neighboring wavelength bins. To model interactions between more widely separated spectral regions, the residual blocks are followed by a multi-head self-attention layer \citep{Vaswani2017}. This layer computes attention weights between spectral positions, while its multiple heads allow different relationships between wavelength regions to be represented in parallel. A global average pooling layer then averages each learned feature channel over the wavelength dimension, producing a fixed-length latent vector \citep{Lin2014}. This representation passes through three fully connected layers, which learn nonlinear combinations of the extracted features, before branching into the task-specific output heads.

The network uses three output heads: (1) a physical parameters head; (2) a primary biosignatures head containing O$_2$ and O$_3$; and (3) an other chemical abundances head containing CH$_4$, CO$_2$, H$_2$O, and N$_2$. The physical head uses a sigmoid activation to constrain its outputs to $[0,1]$, while both chemical heads use a hard-clamp activation (ReLU clipped to $[0,1]$) to enforce non-negative mixing ratios. All three output heads receive the same latent representation from the shared convolutional, attention, and fully connected backbone.

The total training objective is a weighted sum of the per-head MAE losses,
\begin{equation}
\mathcal{L}_{\rm total} = \mathcal{L}_{\rm physical} + 2\,\mathcal{L}_{\rm primary}
+ 1.5\,\mathcal{L}_{\rm other\,chemical}.
\end{equation}
The weights therefore follow the ratio $1{:}2{:}1.5$ for the physical, primary biosignature, and other chemical heads, respectively. The larger weight assigned to the primary-biosignature head was chosen to increase the contribution of O$_2$ and O$_3$ to optimization of the shared network, reflecting their scientific priority in this study. These weights were fixed as a design choice and were not selected through a search over alternative ratios.

All hyperparameters, including filter count, kernel lengths ($h_1$, $h_2$), key dimension ($d_{\mathrm{keys}}$), number of attention heads $q$, dense units $u$, dropout rate $p_{\mathrm{D}}$, and Adam learning rate, were tuned separately for each telescope using Bayesian optimization \citep{omalley2019kerastuner}. We ran 35 trials, training each candidate for 3 epochs to obtain preliminary estimates of the validation loss. The best configuration was subsequently retrained from scratch for 100 epochs on HabEx/SS and 50 epochs on LUVOIR-B, using a batch size of 1024.

\begin{table}
\small
\centering
\caption{Best hyperparameters and training outcomes for HabEx/SS and LUVOIR-B after Bayesian optimization. Validation MAE values are reported on the normalized $[0,1]$ output scale.}
\begin{threeparttable}
\setlength{\tabcolsep}{4pt}
\begin{tabular}{lcc}
\hline
{Parameter} & {HabEx/SS} & {LUVOIR-B} \\[2pt]
\hline
Filters                  & 32            & 16 \\
Kernel length $h_1$      & 70                     & 37 \\
Kernel length $h_2$      & 105          & 75 \\
Key dimension $d_{\mathrm{keys}}$      & 8 & 12 \\
Attention heads $q$  & 4   & 4 \\
Dense units $u$ (first dense)\tnote{a} & 64   & 48 \\
Dropout $p_{\mathrm{D}}$ & 0.30  & 0.35 \\
Learning rate                     & $1.77\times10^{-4}$    & 0.001 \\
\hline
Val. loss MAE (total)    & 0.125               & 0.131 \\
Val. loss MAE (physical parameters)     & 0.072              & 0.074 \\
Val. loss MAE (primary biosignatures)      & 0.010           & 0.011 \\
Val. loss MAE (other chemical abundances)      & 0.021                & 0.024 \\
\hline
\end{tabular}
\begin{tablenotes}
\footnotesize
\item[a] Each output head uses $u/2$ units.
\end{tablenotes}
\end{threeparttable}
\label{tab:4}
\end{table}

\begin{table*}
\centering
\caption{Performance metrics on the held-out test set for the HabEx/SS and LUVOIR-B models. All metrics are calculated in the transformed model-output space used for training. MAE and RMSE are therefore dimensionless, while MAPE and WAPE describe relative errors in that transformed space.}
\label{tab:model_metrics}
\begin{tabular}{lccccc|ccccc}
\toprule
\multirow{2}{*}{{Parameter}} & \multicolumn{5}{c|}{{HabEx/SS}} & \multicolumn{5}{c}{{LUVOIR-B}} \\
& $R^2$ & MAE    & RMSE   & MAPE (\%) & WAPE (\%) & $R^2$ & MAE    & RMSE   & MAPE (\%) & WAPE (\%) \\ \hline
\textit{Physical Parameters}       &       &        &        &           &           &       &        &        &           &           \\
Radius                             & 0.872 & {0.0620} & {0.0832} & {11.9}  & {9.49}  & {0.849} & {0.0692} & {0.0904} & {13.4}  & {10.6}  \\
Gravity                            & {0.872} & {0.0628} & {0.0847} & {13.7}  & {10.6}  & {0.850} & {0.0697} & {0.0915} & {15.2}  & {11.7}  \\
Surface temperature                & {0.485} & {0.0982} & {0.1277} & {361}   & {42.5}  & {0.621} & {0.0817} & {0.1095} & {245}   & {35.3}  \\
Surface Pressure                   & {0.862} & {0.0684} & {0.0939} & {17.9}  & {13.6}  & {0.840} & {0.0759} & {0.1011} & {20.0}  & {15.1}  \\ \hline
\textit{Primary Biosignatures}     &       &        &        &           &           &       &        &        &           &           \\
O$_2$                              & {0.986} & {0.0174} & {0.0299} & {12.8}  & {7.75}  & {0.986} & {0.0174} & {0.0295} & {12.3}  & {7.76}  \\
O$_3$                              & {0.999} & {0.0028} & {0.0045} & {1.13}  & {1.11}  & {0.999} & {0.0032} & {0.0052} & {1.31}  & {1.27}  \\ \hline
\textit{Other Chemical Abundances} &       &        &        &           &           &       &        &        &           &           \\
CH$_4$                             & {0.982} & {0.0138} & {0.0196} & {5.13}  & {3.53}  & {0.981} & {0.0145} & {0.0202} & {5.29}  & {3.72}  \\
CO$_2$                             & {0.956} & {0.0210} & {0.0292} & {6.81}  & {5.59}  & {0.940} & {0.0259} & {0.0341} & {7.92}  & {6.89}  \\
H$_2$O                             & {0.968} & {0.0123} & {0.0166} & {3.54}  & {3.46}  & {0.967} & {0.0127} & {0.0167} & {3.70}  & {3.58}  \\
N$_2$                              & {0.959} & {0.0378} & {0.0600} & {19.2}  & {5.90}  & {0.952} & {0.0441} & {0.0652} & {20.0}  & {6.88}  \\ \hline
\end{tabular}
\end{table*}

During training, we used two callbacks: \texttt{TensorBoard} to monitor diagnostics in real time, and \texttt{EarlyStopping} configured to halt training after 10 epochs without improvement in validation loss. On an NVIDIA A100 GPU, Bayesian hyperparameter optimization took approximately 30 minutes per telescope, and training the final selected model took less than 10 minutes. The LUVOIR-B model contains 92,210 parameters, while the HabEx/SS model contains 494,410. Our complete implementation is available on GitHub\footnote{\url{https://github.com/SarahBarbosa/geexhp}}. The final hyperparameters and validation performance are listed in Table \ref{tab:4}.

\section{Results and discussion} \label{sec:4}

The following analysis examines the model's performance on the test dataset, which contains planetary configurations unseen during the training and optimization process. 

\subsection{Predicted versus true values}

\begin{figure*}
\centering
\includegraphics[width=0.9\textwidth]{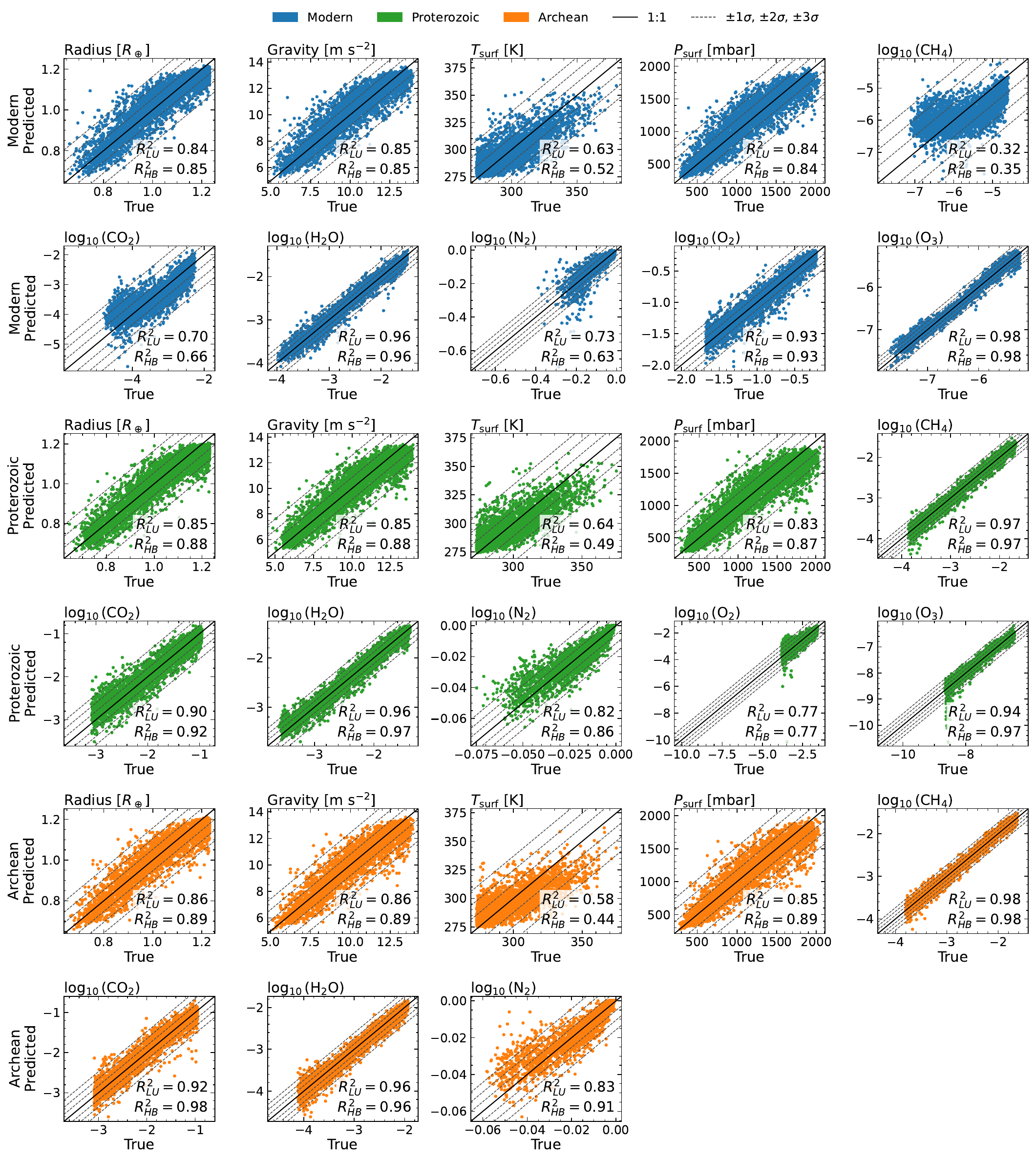}
\caption{Scatter plots illustrating the relationship between predicted and ground-truth values are organized by parameter (columns) and geological era (rows), corresponding to the Modern (blue), Proterozoic (green), and Archean (orange). The solid line denotes the 1:1 relation, while the dashed lines represent deviations of $\pm 1\sigma$, $\pm 2\sigma$, and $\pm 3\sigma$ based on the residuals. $R^2$ is reported for LUVOIR-B ($R^2_{\mathrm{LU}}$) and HabEx/SS ($R^2_{\mathrm{HB}}$) in each panel. The physical parameters are shown in their native units and the positive chemical abundances in $\log_{10}$ mixing-ratio space. Archean O$_2$ and O$_3$ panels are omitted because both abundances are identically zero in that subset.}
\label{fig:predtrue}
\end{figure*}

\begin{figure*}
\centering
\includegraphics[width=0.9\textwidth]{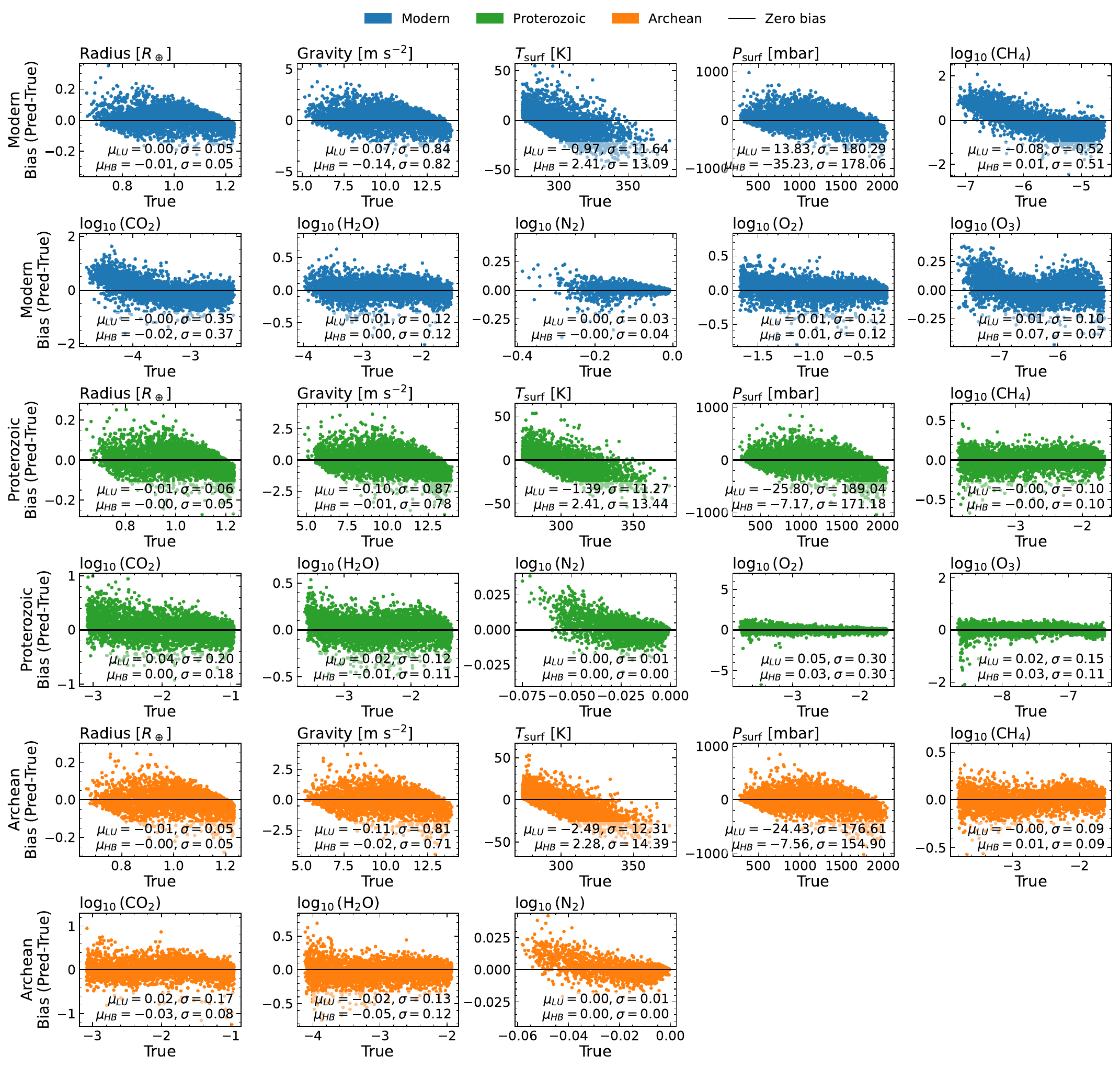}
\caption{Same panel layout and color coding as Fig. \ref{fig:predtrue}. Each panel now shows the bias (predicted minus true) as a function of the true value. The solid horizontal line marks zero bias. The mean bias $\mu$ and standard deviation $\sigma$ of the residuals are quoted for LUVOIR-B and HabEx/SS in each panel. As in Fig. \ref{fig:predtrue}, the Archean O$_2$ and O$_3$ panels are omitted because their logarithmic abundances are undefined.}
\label{fig:bias}
\end{figure*}

Table \ref{tab:model_metrics} reports the aggregate $R^2$, MAE, RMSE, MAPE, and WAPE for each parameter over the held-out test set; full definitions are provided in Appendix \ref{app:metrics_explanation}. In brief, $R^2$ measures predictive performance relative to a mean-value baseline, with $R^2=1$ indicating perfect predictions and $R^2<0$ indicating performance worse than that baseline. MAE is the mean absolute residual, whereas RMSE gives greater weight to large residuals. MAPE averages the absolute relative error of individual predictions and is computed only for non-zero target values, making it sensitive to small denominators. WAPE is the ratio of the summed absolute error to the summed absolute target magnitude and is therefore less sensitive than MAPE to individual near-zero values. All five metrics are calculated in the transformed model-output space used for training. Figures \ref{fig:predtrue} and \ref{fig:bias} present, respectively, the predicted-versus-true relations and residual distributions for the retrieved parameters across the Modern, Proterozoic, and Archean subsets for both LUVOIR-B and HabEx/SS. 

The retrievals of planetary radius, surface gravity, and surface pressure perform well and show only a mild dependence on geological era. In Fig. \ref{fig:predtrue} (columns 1, 2, and 4), the distributions for all three eras lie close to the 1:1 line, with only moderate scatter that largely remains within the 3$\sigma$ bounds (dashed lines). {Quantitatively, the per-era $R^2$ values range from 0.833 to 0.893 across these three quantities and the two instruments.} This accuracy is corroborated by the residual distributions (Fig. \ref{fig:bias}), which are approximately symmetric about zero. The average offsets are constrained to $|\mu|\leq0.01\,R_\oplus$ for radius, $|\mu|\leq0.14$ m s$^{-2}$ for gravity, and $|\mu|\leq36$ mbar for surface pressure. The corresponding dispersions span $0.046$-$0.056\,R_\oplus$, $0.712$-$0.865$ m s$^{-2}$, and $155$-$189$ mbar. Thus, the largest mean offset is no more than approximately 20\% of the corresponding residual dispersion. The era- and instrument-dependent offsets are therefore small in practical terms relative to the point-to-point scatter, and we do not assign them to a specific instrumental mechanism.

The three quantities are also linked by the mass-radius and surface-pressure relations imposed during data generation. Their similar performance therefore does not represent three fully independent spectral constraints, as the network can exploit both spectral information and correlations encoded in the training distribution. Although the Archean subsets generally yield slightly higher $R^2$ values, the differences are modest and are not strictly monotonic for every parameter. We therefore avoid attributing them solely to changes in atmospheric composition.

Surface temperature, on the other hand, emerges as the most challenging physical parameter across all scenarios considered. {Reflected-light spectra contain no direct measurement of planetary thermal emission; temperature is instead encoded indirectly through temperature-dependent absorption, atmospheric scale height, and correlations with the adopted P-T profiles and other planetary properties.} The low intrinsic variance of $T_{\text{surf}}$ in the dataset, already constrained by the liquid-water habitability criterion, further intensifies this limitation. The statistical metrics directly reflect this structural limitation, with $R^2$ values falling between {0.439 and 0.645}, substantially lower than those achieved for other atmospheric parameters. The two instruments show small biases in opposite directions: LUVOIR-B underestimates $T_{\rm surf}$ by approximately 1.0-2.5 K, whereas HabEx/SS overestimates it by approximately 2.3-2.4 K. These offsets are small relative to the residual dispersions of 11.3-14.4 K. The dominant behavior is therefore a regression toward the center of the sampled temperature range. This compression manifests as the fan-shaped pattern visible in the corresponding panels of Fig. \ref{fig:predtrue}.

Water vapor is one of the most consistently well-retrieved species in every epoch, with $R^2$ between 0.956 and 0.966 across the six era-instrument combinations. The ubiquity of H$_2$O absorption signatures across many wavelength regions introduces redundancy in the diagnostic information, which helps to stabilize the retrievals against moderate noise and degeneracies with other parameters. Methane exhibits a similar sensitivity to abundance regime. In the Proterozoic and Archean atmospheres, where CH$_4$ is plentiful, its strong NIR bands yield near-ideal performance with $R^2 \approx 0.97$-$0.98$ for both models. In contrast, in the Modern atmosphere, where methane is only a trace gas, weaker absorption features diminish the available information, and the performance correspondingly drops to $R^2=0.317$ for LUVOIR-B and $0.348$ for HabEx/SS. The corresponding residual dispersions exceed 0.5 dex, indicating that Modern-Earth CH$_4$ abundances are not reliably recovered in the present setup.

CO$_2$ follows a similar abundance-dependent trend, but its performance varies more strongly across eras and instruments. In the Modern subset, $R^2$ is 0.702 for LUVOIR-B and 0.664 for HabEx/SS. It increases to 0.896 and 0.917, respectively, in the Proterozoic and to 0.918 and 0.983 in the Archean. The stronger performance outside the Modern subset is consistent with the larger CO$_2$ abundances and greater spectral leverage in those training distributions, although differences in their within-era target ranges also contribute to the changes in $R^2$.

Ozone {is among the most consistently constrained chemical species. In the Modern subset, $R^2$ reaches 0.977 for LUVOIR-B and 0.975 for HabEx/SS, consistent with the strong O$_3$ absorption in the Hartley-Huggins UV region. In the Proterozoic, LUVOIR-B decreases to $R^2=0.942$, with the residual dispersion increasing from 0.097 to 0.147 dex. HabEx/SS changes more modestly, from $R^2=0.975$ to $0.967$, with $\sigma$ increasing from 0.074 to 0.108 dex. Thus, the Proterozoic results show a broader and instrument-dependent O$_3$ retrieval}. This echoes the conclusions of \citet{Damiano2023}, who demonstrated using ExoReL Bayesian retrievals that UV coverage is a decisive discriminant for Proterozoic-era O$_3$ detection. This result underscores the importance of UV sensitivity in HWO bandpass trade studies, particularly for terrestrial targets that may be in intermediate oxidation states. 

O$_2$ displays a similar trend, with strong Modern-era correlations giving {$R^2=0.929$ for LUVOIR-B and $0.926$ for HabEx/SS}, whereas the low-O$_2$ Proterozoic interval experiences a moderate decline to {$R^2\simeq0.77$ for both instruments}. For all species, the mean residuals remain bounded by $|\mu|<0.08$ dex, with the largest absolute value being 0.077 dex for Modern CH$_4$ with LUVOIR-B. These modest global offsets do not exclude abundance-dependent errors or effects introduced by the training priors. For example, the HabEx/SS Modern O$_3$ residual has $\mu=0.067$ dex and $\sigma=0.074$ dex, revealing a small but coherent positive displacement despite its high $R^2$.

Among the retrieved species, N$_2$ deserves particular attention. Being a homonuclear diatomic molecule, N$_2$ has no permanent electric dipole moment and therefore lacks strong, isolated monomer absorption bands over the wavelength range considered here. Its spectral constraint is instead indirect, arising from effects such as Rayleigh scattering and correlations with surface pressure, mean molecular weight, and the abundances of spectrally active gases \citep{Schwieterman2015}. The compositional-closure condition and the geological prior imposed during data generation provide additional predictive correlations. This distinction is especially important when comparing eras. In the Modern subset, $R^2$ is 0.732 for LUVOIR-B and 0.633 for HabEx/SS, compared with 0.833 and 0.912 in the Archean. However, Archean N$_2$ occupies a much narrower abundance interval and has residual dispersions of only 0.005 and 0.004 dex. Its higher $R^2$ therefore reflects the combined effects of the epoch-dependent distribution, compositional closure, and indirect continuum information.

\citet{Hall2023} showed that reflected-light retrievals can identify N$_2$-dominated backgrounds at S/N$\sim20$ when the spectrum extends beyond approximately $1.6\,\mu$m, but also found a residual O$_2$-N$_2$ degeneracy that is fully excluded only at S/N$\sim40$. They further showed that N$_2$ and CO backgrounds cannot be distinguished with only $0.2$-$1.1\,\mu$m coverage. Although both instrument concepts considered here nominally extend beyond $1.6\,\mu$m, our population contains CO only at the trace abundances generated around the geological templates, does not include CO-rich or CO-dominated backgrounds, and does not retrieve CO. Our reported N$_2$ performance therefore does not test the CO-N$_2$ degeneracy or establish a unique identification of N$_2$ under broader atmospheric priors.

\subsubsection{Behavior at zero true mixing ratios}

A particularly relevant fraction of the test set represents Archean-like planets ({3,353} cases) that, by design, contain no O$_2$ or O$_3$ (Table \ref{tab:2}). For any biosignature retrieval framework, being able to correctly infer the absence of a biosignature gas is just as critical as detecting it when it is present; any recurring tendency toward false positives would directly affect how HWO targets are selected and prioritized.

Non-negativity is enforced by the hard-clamp activation in the biosignature output head (Section \ref{sec:3}), which constrains predictions to $[0,1]$ and drives outputs toward zero when the spectral signal is weak. However, the zero-abundance case is intrinsically special in our label representation. Molecular abundances are mapped into model-output space through a species-dependent power transform (Appendix \ref{app:A}), but truly absent species remain exactly at zero under this transformation. As a result, the network is not learning a smoothly varying target near zero for Archean O$_2$ and O$_3$, but rather a distribution with a hard boundary at zero and only one allowed sign of deviation.

For O$_2$, HabEx/SS returns strictly zero predictions in all 3,353 cases, whereas LUVOIR-B yields a single non-zero output, reaching $-3.60$ dex. For O$_3$, HabEx/SS produces seven non-zero outputs and LUVOIR-B produces six, but these are overwhelmingly concentrated at extremely small values, approaching the numerical floor of the output representation: they span $1.4\times10^{-45}$ to $2.0\times10^{-21}$, the lower end being exactly the smallest representable single-precision value. In other words, ozone shows more frequent departures from an exact zero prediction than oxygen, a consequence of the zero-boundary leakage mechanism described above, but these departures typically occur at abundances so low that they are not physically distinguishable from a null result.

Accordingly, these non-zero outputs do not constitute detections. In a retrieval context, a positive inference requires the predicted abundance to be meaningfully separated from zero relative to its uncertainty. By that criterion, {all} of the O$_3$ residual outputs in Archean cases are effectively consistent with a null result. The O$_2$ case is somewhat cleaner in the sense that exact zeros are recovered more often, although the rare non-zero excursions can reach larger amplitudes. HabEx/SS is the more conservative of the two for O$_2$, returning no non-zero outputs at all, while for O$_3$ the two instruments are indistinguishable.

Overall, the false-positive rate for non-zero O$_2$ or O$_3$ in truly anoxic atmospheres remains below $\sim$ {0.2}\% of Archean test cases for both models, and below {0.03}\% for O$_2$ specifically. 

\subsection{Feature importance via Integrated Gradients}\label{sec:4.2}

To understand which spectral features {contribute most strongly to the chemical predictions of our model}, we employed the Integrated Gradients (IG) technique, an eXplainable AI (XAI) method {introduced} by \cite{Sundararajan2017}. This method attributes the prediction for each gas to individual input features (i.e., wavelengths) by {integrating} the model gradients along a linear path from a baseline to the input spectrum. For our analysis, the baseline is a vector of zeros in the standardized input space (equation \ref{a1}), which corresponds to a flat spectrum at the training-set mean albedo of each instrument channel (hereafter, the channel-mean baseline). For gas $g$, the resulting signed attribution, $\mathrm{IG}_g(\lambda)$, quantifies the contribution of each spectral bin to the transformed network output relative to this baseline.

\begin{figure*}
\centering
\includegraphics[width=\textwidth]{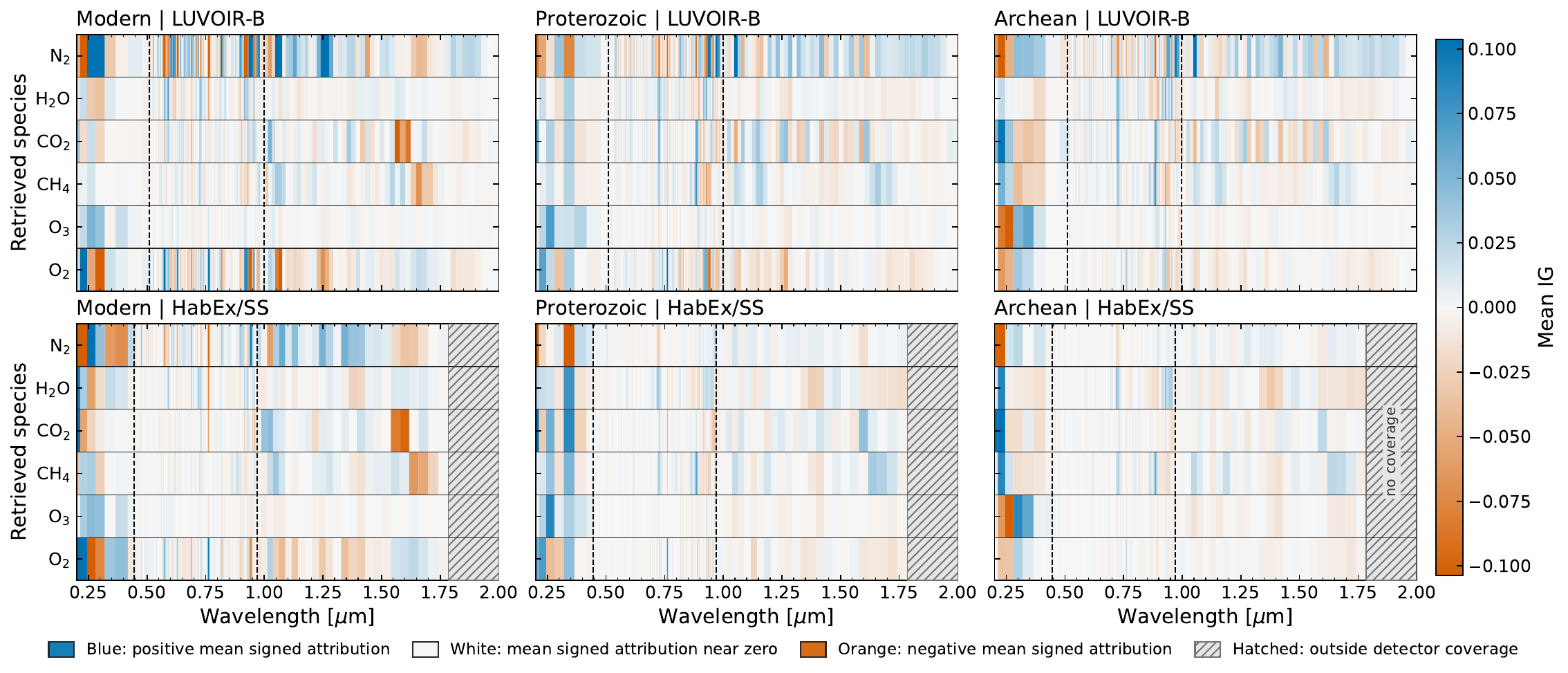}
\caption{Mean signed Integrated-Gradients (IG) attribution for the six chemical outputs. Rows of panels correspond to LUVOIR-B (top) and HabEx/SS (bottom), while columns correspond to the Modern, Proterozoic, and Archean eras. Within each panel, wavelength is shown on the horizontal axis and each chemical species occupies a separate horizontal strip on the vertical axis. Each cell gives the signed IG attribution for the species in that row, averaged over all available test spectra in the corresponding era. Positive attribution relative to the channel-mean baseline is shown in blue, negative attribution in orange, and a near-zero signed mean in white. Cell widths follow the actual wavelength-bin widths, and thin horizontal lines separate adjacent species. Symmetric color limits are shared by both instruments and set by the 99th percentile of the absolute IG values pooled across both telescopes, all species, and all eras. Dashed vertical lines mark the UV/VIS and VIS/NIR boundaries. Hatched regions lie outside the HabEx/SS wavelength coverage, which ends near $1.8\,\mu\mathrm{m}$.}
\label{fig:ig_heatmap}
\end{figure*}

Instead of calculating IG for a single representative example from each era, we aggregate the attributions over the full test subset. IG is calculated separately for each of the six chemical outputs using 24 integration steps. For a selected gas, the predictions for that gas are summed across the independent spectra in each computational batch and differentiated with respect to the corresponding inputs. Because the network processes each spectrum independently, this operation returns a separate attribution curve for every spectrum while allowing the gradients to be evaluated efficiently in parallel. We then average the signed attribution separately within each geological era, using all available test spectra. The resulting population-averaged signed attribution maps are shown in Fig. \ref{fig:ig_heatmap}. Signed averaging preserves the population-mean direction of the attribution, but positive and negative sample-level contributions can cancel. Consequently, a near-zero mean attribution does not necessarily imply that a wavelength is unimportant for every individual spectrum. The IG values are expressed in the transformed model-output space.

{To summarize the wavelength distribution of attribution independently of its sign, we project the absolute IG values onto the wavelength axis after normalization within each valid sample-output pair, and divide by the width of each bin so that the result is an attribution density per unit wavelength (Fig.~\ref{fig:ig_projected}).} The full definition and implementation of this procedure are given in Appendix~\ref{app:ig_projection}. Because the curve integrates to unity over wavelength, the area it encloses within each channel is the fraction of the total attribution carried by that channel. The VIS carries between $58$ and $71$\% across the three eras and both instruments, the NIR between $16$ and $23$\%, and the UV between $6.9$ and $13.1$\% for LUVOIR-B and $19.3$ and $26.2$\% for HabEx/SS. In LUVOIR-B the NIR therefore carries one and a half to three times the integrated attribution of the UV. The strongest feature of the projected curve is the O$_2$ A band at $0.76\,\mu$m, which is the global maximum in five of the six instrument-era combinations, with further resolved structure near $0.58$ and $0.73\,\mu$m in both instruments and near $0.63$ and $0.68\,\mu$m in LUVOIR-B. The NIR maxima are led by a feature near $1.05\,\mu$m, with weaker structure near $1.57$-$1.60\,\mu$m. Although the HabEx/SS curves reach mean UV densities two and a half to three and a half times larger than those of LUVOIR-B, the VIS carries the larger integrated fraction in every case and the higher mean density per unit wavelength in every case, although only narrowly in Archean HabEx/SS ($1.10$ against $1.05\,\mu$m$^{-1}$).

\begin{figure}
\centering
\includegraphics[width=\columnwidth]{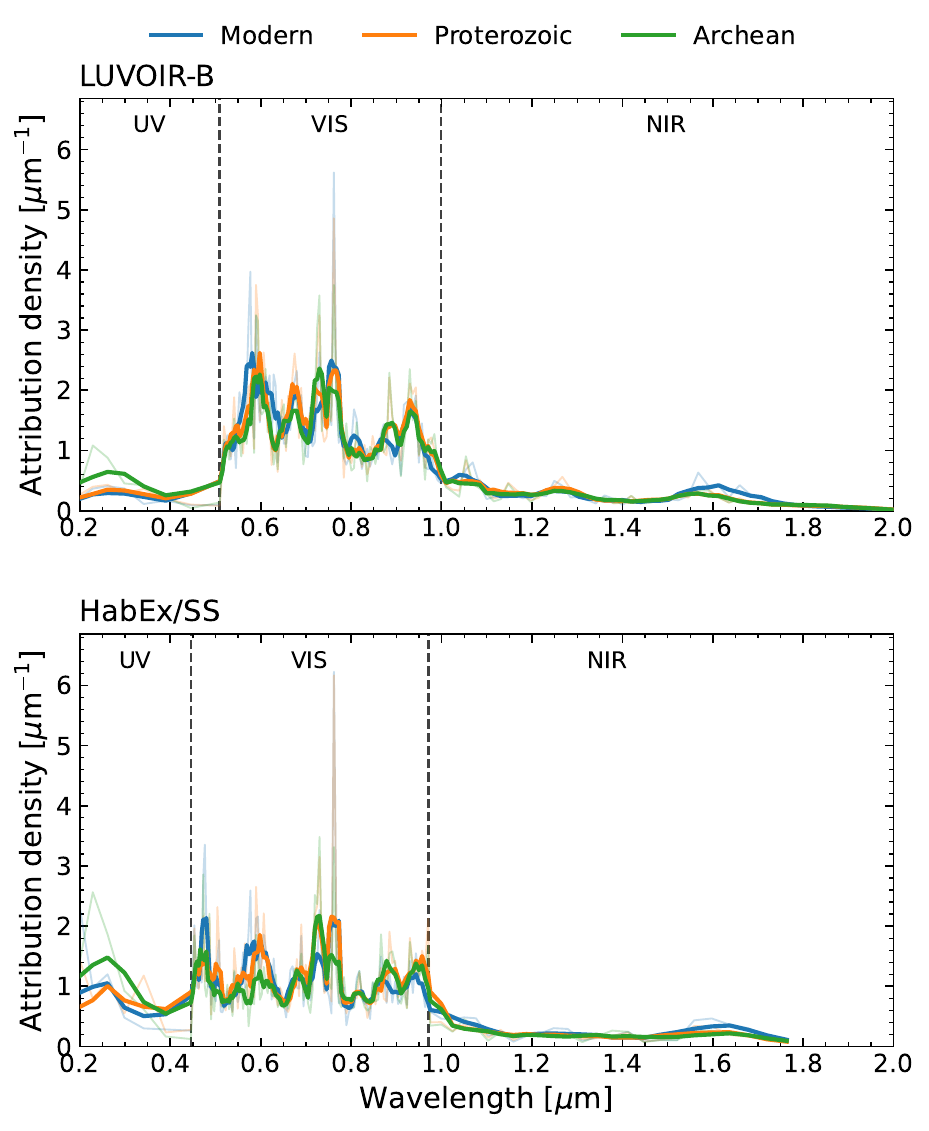}
\caption{Projected integrated-gradient attribution spectra for LUVOIR-B (top panel) and HabEx/SS (bottom panel). For each era, the curve is obtained by taking the absolute IG for every valid sample-output pair, normalizing its bin weights to unit total attribution, dividing those weights by the corresponding wavelength-bin widths, and then averaging over samples and outputs. The plotted quantity is therefore an attribution density whose integral over wavelength is unity. Channel fractions are calculated from the normalized bin weights before conversion to density. Thin lines show the unsmoothed projected scores; thick lines show a light moving-average smoothing for display only. Dashed vertical lines mark the UV/VIS and VIS/NIR channel boundaries.}
\label{fig:ig_projected}
\end{figure}

Figs. \ref{fig:ig_heatmap} and \ref{fig:ig_projected} answer different questions. The heatmap reports the attribution assigned to each detector channel, so a wide UV bin accumulates the attribution of a broad interval into a single value; the projected curve divides by the bin width and therefore measures attribution per unit wavelength. Because the UV channels are about eight times wider than the visible ones, a species whose dominant \emph{channel} lies in the UV can still have its highest attribution \emph{density} in the visible. The two views are complementary: the first identifies which channels a detector must provide, the second where observing effort is most productively distributed in wavelength.

At the species-resolved level, Fig. \ref{fig:ig_heatmap} shows the population-mean direction of the attribution separately for each chemical output. In the Modern cases, O$_3$ displays pronounced signed UV structure for both instruments, coinciding with the Hartley-Huggins region. O$_2$ shows attribution across the UV and VIS, together with additional signed structure near the VIS-NIR boundary. CO$_2$ and CH$_4$ exhibit more localized contributions in the NIR, particularly around $1.5{-}1.7\,\mu\mathrm{m}$, while H$_2$O shows weaker, more distributed structure. N$_2$ has mixed positive and negative attribution at short wavelengths, consistent with the network using continuum- and scattering-sensitive information.

Read per detector channel, the species-resolved normalized absolute-attribution profiles recover several well-established molecular regions even though their locations were not supplied explicitly to the network. In the Modern subset, the largest O$_2$ attribution occurs at 762 nm for both instruments, coinciding with the O$_2$ A band. O$_3$ peaks within the Hartley band, at 261 nm for LUVOIR-B and 299 nm for HabEx/SS, while CH$_4$ peaks at 1659 nm for LUVOIR-B and 1638 nm for HabEx/SS, and CO$_2$ at 1568 and 1598 nm, respectively. These locations lie within one or two instrumental bins of spectral regions commonly examined in HWO-relevant reflected-light studies: approximately 0.76 $\mu$m for O$_2$, 0.20-0.35 $\mu$m for the O$_3$ Hartley-Huggins system, 1.65 $\mu$m for CH$_4$, and 1.57-1.60 $\mu$m for CO$_2$ \citep{brandt2014,Damiano2023,young2024,Tokadjian2024}.

At the task-aggregated level, the attribution density falls to a comparatively small tail beyond approximately $1.7\,\mu$m, after the principal NIR maxima of CO$_2$ and CH$_4$. This identifies $\sim1.7\,\mu$m as a practical point of diminishing aggregate retrieval leverage for the atmospheric population considered here. The result agrees with independent studies that identify $1.7\,\mu$m as necessary for biosignature contextualization \citep{KrissansenTotton2025}, obtain an optimal CO$_2$-motivated cutoff of $1.68\,\mu$m \citep{Hagee2026}, or find that CH$_4$ remains detectable under truncation at $1.6$-$1.7\,\mu$m \citep{Tokadjian2024}. This agreement is particularly notable because those studies rely on forward-model and Bayesian retrieval analyses, while our results are based on neural-network attribution.

The Modern-Earth O$_3$ attribution also contains a broad visible contribution, with a resolved local maximum near $0.6\,\mu$m coincident with the Chappuis band. To quantify this species-specific contribution, we use the same per-sample normalization of the absolute attribution defined in equation~\ref{eq:ig_norm_per_output} and underlying Fig.~\ref{fig:ig_projected}, while keeping the O$_3$ output separate from the average over the six chemical outputs. Under this definition, the Chappuis region ($0.40$-$0.70\,\mu$m) carries $34.3\%$ of the total Modern-Earth O$_3$ attribution for LUVOIR-B and $37.8\%$ for HabEx/SS, against $12.1\%$ and $17.8\%$ for the Hartley--Huggins region below $0.35\,\mu$m.

The Chappuis contribution is less visually prominent in the population-mean signed attribution of Fig.~\ref{fig:ig_heatmap} because it is spread across many narrow VIS bins and because its sign varies among test spectra, so that the population average partially cancels. Measuring this as $\sum_j |\langle \mathrm{IG}_j \rangle| / \sum_j \langle |\mathrm{IG}_j| \rangle$ within each region, the Modern-Earth O$_3$ attribution is almost entirely sign-coherent across the Hartley--Huggins band ($0.99$ for LUVOIR-B, $0.82$ for HabEx/SS) but only partially so across the Chappuis band ($0.36$ and $0.38$). The two regions therefore provide complementary leverage. At Modern-Earth abundance the broad Chappuis band carries the larger integrated absolute attribution, whereas the Hartley-Huggins region provides the more localized and sign-coherent signal, and its share grows as O$_3$ decreases, rising to $13.6\%$ and $23.1\%$ in the Proterozoic subset.

The Proterozoic panels retain clear short-wavelength O$_3$ attribution for both instruments. O$_2$ becomes weaker and less structured, while CH$_4$ and CO$_2$ retain additional signed structure in the NIR.

In the Archean subsets, the short-wavelength contribution increases substantially but does not dominate the integrated attribution budget for most outputs. The fraction of attribution below $0.35\,\mu\mathrm{m}$ rises from 0.12 in the Modern subset to 0.31 in the Archean for LUVOIR-B O$_3$, and from 0.18 to 0.44 for HabEx/SS. For HabEx/SS O$_3$, the short-UV contribution therefore becomes comparable to that of the VIS. For every other output, however, the VIS still contains the majority of the integrated attribution, with fractions between 0.55 and 0.84. The short-wavelength continuum of the anoxic, CO$_2$- and CH$_4$-enhanced Archean templates is therefore an important source of learned leverage, but it does not replace the VIS as the dominant channel when attribution is integrated over wavelength.

The Archean O$_2$ and O$_3$ rows also require a different attention. Although both gases are identically zero in the Archean targets, neither population-mean row is entirely featureless. In the species-resolved normalized profiles, the largest O$_2$ attribution remains near the 762-nm A band and the O$_3$ maximum remains near 261 nm. However, the location of the largest bin must be considered together with the much smaller overall attribution amplitude. For O$_2$, the population-mean signed sum, $\langle\sum_\lambda \mathrm{IG}_{\mathrm{O_2}}(\lambda)\rangle$, changes from $+0.14$ in the Modern subset to $-0.32$ in the Archean for LUVOIR-B, while the median total absolute attribution decreases by factors of 4.2 and 8.2 for LUVOIR-B and HabEx/SS, respectively. These values are expressed in the transformed network-output space. The negative signed sum indicates that the observed Archean spectrum lowers the transformed O$_2$ output relative to the channel-mean baseline, while the reduction in absolute attribution shows that the output is much less spectrally responsive in this regime.

This behavior is consistent with the zero-abundance analysis, in which the models return an exactly zero O$_2$ abundance for 99.97\%-100\% of the Archean spectra. Nevertheless, IG relative to the channel-mean baseline does not demonstrate that the model would respond correctly to the counterfactual insertion of an O$_2$ or O$_3$ absorption band. It shows only which observed bins influence the output along the selected baseline-to-spectrum path.

\subsection{Uncertainty estimation}\label{sec:uncertainty}

We decompose the predictive uncertainty into two complementary components associated with observational noise and model ambiguity. The data (aleatoric) uncertainty, $\sigma_{\text{data}}$, is estimated by {parametric observational-noise resampling of the input spectrum}: for each test case, we generate $B=5000$ independent Gaussian noise realizations using the noiseless apparent-albedo spectrum and the per-wavelength noise vector stored in the test set, and pass each realization through the network with dropout disabled. {The standard deviation of the resulting prediction ensemble quantifies how fluctuations under the adopted observational-noise model propagate into the retrieved parameters, with dropout disabled so that stochastic variation from the dropout masks is excluded.}

The model (epistemic) uncertainty, $\sigma_{\text{model}}$, is estimated using Monte Carlo (MC) Dropout \citep{Gal2015}. In this case, dropout layers remain active at inference time, and $N_{\text{MC}}=5000$ stochastic forward passes are performed for a fixed noise realization. The spread of the resulting predictions measures the sensitivity of the retrieval to the particular subnetwork activated in each draw, and therefore provides an approximate estimate of the uncertainty associated with the learned model. We emphasize, however, that this estimate is not expected to be fully equivalent to a rigorously calibrated Bayesian posterior. In particular, SpatialDropout1D acts on convolutional feature maps instead of individual weights, and the dropout rates adopted here were selected through Bayesian hyperparameter optimization based on predictive performance and training stability; posterior calibration was not part of the optimization criterion. As a result, MC Dropout may underestimate the true epistemic variance \citep[e.g.,][]{Folgoc2021}. The observational-noise resampling provides a complementary estimate obtained by varying the adopted noise realization while holding the dropout masks inactive.

Because these two components are measured under independent perturbation schemes, we combine them in quadrature to define a total predictive uncertainty,
\begin{equation}\label{eq:sigma_total}
\sigma_{\rm total} = \sqrt{\sigma_{\rm model}^{2} + \sigma_{\rm data}^{2}}.
\end{equation}
All uncertainty quantities are expressed in the normalized model-output space in which the network is trained, so that comparisons remain meaningful across parameters with different physical units and dynamic ranges.

We emphasize that this uncertainty framework is intended primarily as a rapid comparative tool for retrieval screening, target triage, sensitivity analyses, and instrument-level performance assessment. In that context, limitations such as mild undercoverage at the $1\sigma$ level affect the interpretation of the narrowest uncertainty intervals, but do not preclude $\sigma_{\rm total}$ from providing useful information on relative recoverability, on the separation between data-limited and model-limited regimes, and on how retrieval behavior changes across atmospheric scenarios and observing assumptions. Full implementation details for both the bootstrap and MC Dropout procedures are given in Appendix~\ref{app:uncertainty}. Appendix \ref{app:cornerplot} complements this with per-sample {joint predictive distributions} for representative Modern, Proterozoic, and Archean test spectra, showing how the two uncertainty components manifest at the level of individual retrievals. 

\subsubsection{Relative importance of data and model uncertainty}\label{sec:4.3.1}

A key issue within this operational decomposition is whether the estimated predictive variance at the nominal noise level is associated primarily with observational-noise propagation or with dropout-induced model variability. When $\sigma_{\text{data}} \gg \sigma_{\text{model}}$, reducing the adopted noise scale is expected to reduce a substantial component of $\sigma_{\rm total}$. Conversely, when $\sigma_{\text{model}} \gg \sigma_{\text{data}}$, reducing observational noise alone has less leverage on the quadrature estimate.

Fig. \ref{fig:uncerfra} shows the median fraction of total variance attributable to data uncertainty, $\sigma_{\rm data}^2 / \sigma_{\rm total}^2$, for each of the ten retrieved parameters, evaluated at the default noise scaling $s = 1.0$ (corresponding to the nominal mission noise level; see Appendix~\ref{app:uncertainty} for the definition of $s$). {At this noise level, the median summaries differ between the two instrument concepts. For LUVOIR-B, the data fraction lies below 50\% for six of the ten parameters; the four medians above the line are only slightly larger than 0.5, and none exceeds approximately 0.56. The LUVOIR-B budget is therefore broadly mixed, with a model-dominated median for most outputs. For HabEx/SS, eight of the ten medians exceed 50\%, reaching approximately 0.70 for $T_{\rm surf}$; only O$_3$ and CH$_4$ remain below the line, at approximately 0.11 and 0.45, respectively.

O$_3$ has the smallest median data fraction in both instruments, with $\sigma_{\rm data}^2/\sigma_{\rm total}^2 \approx 0.07$ for LUVOIR-B and $0.11$ for HabEx/SS. This pattern is consistent with the strong and spectrally isolated Hartley-Huggins absorption limiting the response of the prediction to the imposed noise perturbations; the fraction itself, however, does not identify the origin of the MC-Dropout spread. For LUVOIR-B, CH$_4$ and H$_2$O also have model-dominated median estimates, with data fractions of approximately 0.31 and 0.37, respectively, although the H$_2$O interquartile range extends across the 50\% reference.

\begin{figure}
\centering
\includegraphics[width=\columnwidth]{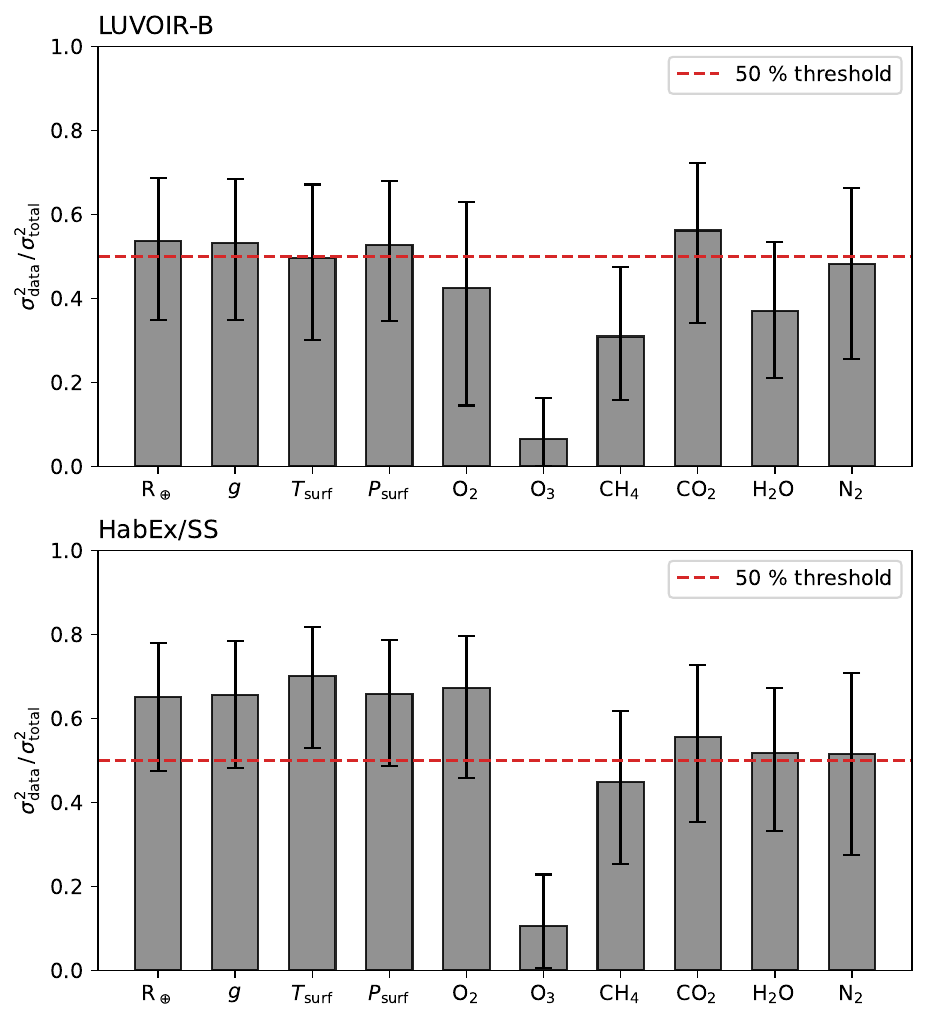}
\caption{Fraction of total variance attributable to data (aleatoric) uncertainty, $\sigma^2_{\rm data}/\sigma^2_{\rm total}$, for each retrieved parameter {at the nominal noise scaling $s=1$}, for LUVOIR-B (top panel) and HabEx/SS (bottom panel). Bars indicate median values over the test set, and error bars span the 25th-75th percentile range. The dashed horizontal line marks the 50\% threshold.}
\label{fig:uncerfra}
\end{figure}

Within LUVOIR-B, the largest median aleatoric contribution occurs for CO$_2$ ($0.56$), followed by $R_\oplus$, $g$, and $P_{\rm surf}$ ($0.53$-$0.54$), while $T_{\rm surf}$ lies close to equal partition at approximately $0.50$. These parameters therefore occupy a mixed regime in which the two estimated components are comparable. For $T_{\rm surf}$, whose data fraction reaches approximately $0.70$ in HabEx/SS, the result shows that observational-noise perturbations make an important contribution to the predictive spread alongside the poor $R^2$ performance reported in Section~\ref{sec:4}. Because this fraction is relative and is computed through the present trained network, it does not establish an irreducible temperature-information limit or determine how a perfectly specified model would perform. For CO$_2$, the comparatively large data fraction may reflect the overlap of its NIR absorption bands with those of CH$_4$ and H$_2$O.

HabEx/SS has a higher median data fraction than LUVOIR-B for nine of the ten retrieved parameters; CO$_2$ is the single exception, at $0.56$ against $0.55$. For the four physical parameters, the absolute differences range from approximately 0.12 to 0.20. This is consistent with a larger relative contribution from observational noise under the adopted HabEx/SS configuration. The two concepts also differ in wavelength coverage, spectral resolution, throughput, and observing architecture, however, so the figure does not isolate collecting area as the cause of the difference. The interquartile ranges show appreciable sample-to-sample variation for most outputs and frequently cross the 50\% reference.

At $s=1$, this operational decomposition therefore suggests different median balances for the two instrument concepts. For HabEx/SS, reducing observational noise has greater potential leverage on the estimated variance of most parameters, particularly $T_{\rm surf}$. For LUVOIR-B, the expected leverage is more parameter-dependent: CO$_2$ and the physical parameters lie closest to equal partition, whereas O$_3$ and CH$_4$ show the clearest dominance of the MC-Dropout component. In particular, the low data fractions of O$_3$ and CH$_4$ do not imply that no realistic exposure strategy can improve their retrievals, but only that, at $s=1$, their dropout-induced variance exceeds the propagated observational-noise variance. The noise-scaling experiments presented in the following subsection provide the more direct test of how the retrieval changes when the observational noise is varied. Overall, the results support a hybrid strategy in which observational improvements are considered for parameters with large data fractions, while improvements to the inference model, training-set coverage, and uncertainty calibration are emphasized for parameters dominated by the MC-Dropout component.

\subsubsection{Response to the noise scale factor}

To assess how retrieval precision scales with observational quality, we repeat the spectral bootstrap at four noise scale factors $s \in \{0.5, 1.0, 2.0, 4.0\}$. Fitting $\sigma_{\rm data}\propto s^{\alpha}$ to the four medians gives exponents between $0.89$ and $1.16$ for LUVOIR-B and between $0.64$ and $1.19$ for HabEx/SS, so the scaling is close to linear for most parameters (Fig. \ref{fig:uncers}). Two departures are worth noting. O$_3$ is mildly super-linear in both instruments ($\alpha\simeq1.16$ and $1.19$), and $T_{\rm surf}$ in HabEx/SS is markedly sub-linear ($\alpha=0.64$): its median $\sigma_{\rm data}$ grows only from $0.110$ to $0.128$ between $s=2$ and $s=4$, indicating that the predicted temperature is already compressed against the bounds of the sampled prior and can no longer respond to further degradation of the spectra. The different slopes in Fig. \ref{fig:uncers} reflect the fact that the retrieved parameters are not encoded equally in the spectrum. Quantities associated with strong, spectrally localized, and weakly degenerate features remain comparatively insensitive to changes in the noise level, whereas parameters constrained only indirectly through continuum shape, pressure effects, or strongly covariant signatures exhibit a steeper increase in uncertainty as the noise scale factor grows.

\begin{figure}
\centering
\includegraphics[width=\columnwidth]{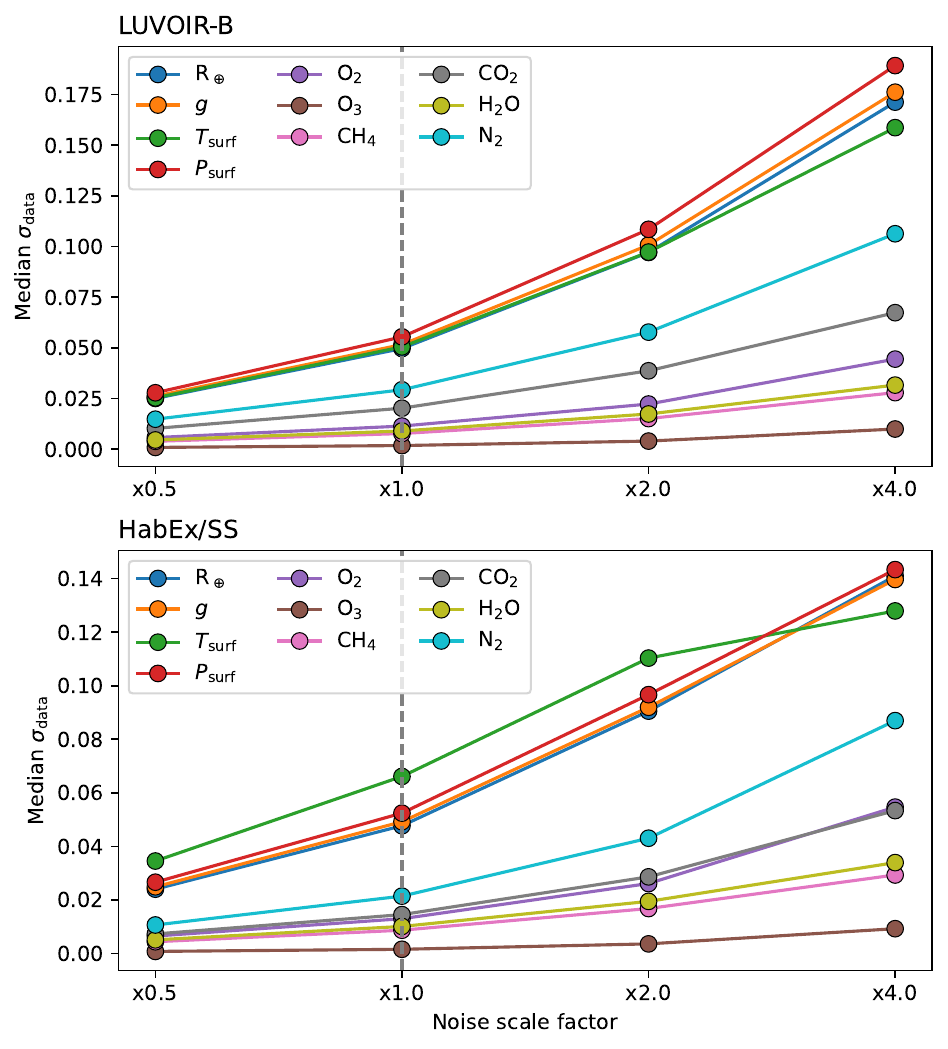}
\caption{Data uncertainty as a function of noise scale factor. Median aleatoric uncertainty, $\sigma_{\rm data}$, estimated via spectral bootstrap after rescaling the PSG per-wavelength noise vector by a factor of $0.5$, $1$, $2$, and $4$ for LUVOIR-B (top panel) and HabEx/SS (bottom panel). The dashed vertical line marks the default noise scale factor ($s = 1.0$). In LUVOIR-B panel, CH$_4$ and H$_2$O are overlaid.}
\label{fig:uncers}
\end{figure}

For the chemically well-constrained species (O$_3$, O$_2$, H$_2$O, CH$_4$), the rise in $\sigma_{\rm data}$ remains modest \emph{in absolute terms}: across the full factor-of-eight range in $s$, their median $\sigma_{\rm data}$ increases by at most $0.048$ in normalized output units, against $0.13$-$0.16$ for the physical parameters in LUVOIR-B and $0.09$-$0.12$ in HabEx/SS. In relative terms the ordering is different, and O$_3$ in fact grows fastest of all, by factors of $11.3$ (LUVOIR-B) and $12.0$ (HabEx/SS) over the same range, simply because it starts from the smallest value. Combined with the variance decomposition in Fig. \ref{fig:uncerfra}, the noise scale at which each parameter reaches equal partition between the two components separates these species clearly. Taking the value of $s$ at which the median data fraction crosses $0.5$, O$_2$ reaches it at $s\simeq1.2$ for LUVOIR-B and $0.7$ for HabEx/SS, H$_2$O at $1.3$ and $1.0$, and CH$_4$ at $1.5$ and $1.1$, whereas O$_3$ requires $s\simeq2.9$ and $2.3$. Only O$_3$ therefore remains firmly model-limited under substantially degraded observing conditions; the other three cross into a data-dominated regime close to, or only slightly above, the nominal noise level. For the fixed frame time adopted here, the PSG source, background, dark-current, and read-noise variances scale linearly with the number of integrations. The noise amplitude therefore scales as $t^{-1/2}$, allowing an exposure-equivalent conversion $t_{\rm eq}=1000\,{\rm h}/s^2$, provided no systematic contrast floor intervenes. Within the present fixed-network decomposition, the equal-variance
crossing times range from 120\,h for LUVOIR-B O$_3$ to 2,464\,h for HabEx/SS $T_{\rm surf}$. At 250\,h ($s=2$), nine of the ten outputs have a larger observational-noise component in both concepts; only O$_3$ remains dominated by the MC-Dropout component.

In absolute terms, the largest increases in $\sigma_{\rm data}$ with noise scale are seen for the physical parameters, while among the chemical species the strongest growth appears for N$_2$ and CO$_2$. This does not imply, however, that all of these parameters are equally close to a data-limited regime. When combined with the variance decomposition in Fig. \ref{fig:uncerfra}, the results indicate that $T_{\rm surf}$ is already near a mixed regime at the default noise level{, crossing equal partition at $s\simeq1.0$ for LUVOIR-B and $s\simeq0.6$ for HabEx/SS}, while CO$_2$ is the clearest chemical case in which increasing noise pushes the retrieval toward aleatoric dominance{: it crosses at $s\simeq0.9$ in both instruments, the earliest of any parameter in LUVOIR-B, although in HabEx/SS O$_2$ crosses earlier still at $s\simeq0.7$}. N$_2$ also becomes substantially more uncertain as $s$ increases, but its interpretation remains more intermediate, since its larger absolute $\sigma_{\rm data}$ does not translate as directly into the highest data fraction.

Fig. \ref{fig:uncers} shows that worsening the noise level does not affect all targets equally. The best-constrained species degrade comparatively slowly in absolute terms, whereas CO$_2$, N$_2$, and the physical parameters accumulate the largest absolute increases in $\sigma_{\rm data}$. The behaviour of $T_{\rm surf}$ is more nuanced: it responds strongly between $s=0.5$ and $s=2$ but then saturates, most visibly in HabEx/SS, so that its uncertainty ceases to track the noise level once the retrieval has lost most of its temperature information. Comparing the two instruments, every parameter except CO$_2$ reaches equal partition at a lower noise scale in HabEx/SS than in LUVOIR-B, with crossing scales spanning $0.6$-$2.3$ against $0.9$-$2.9$. CO$_2$ is the same single exception that appears in Section~\ref{sec:4.3.1}, where it is the only parameter whose LUVOIR-B data fraction exceeds that of HabEx/SS; the two diagnostics therefore agree. 

\begin{figure*}
\centering
\includegraphics[width=\textwidth]{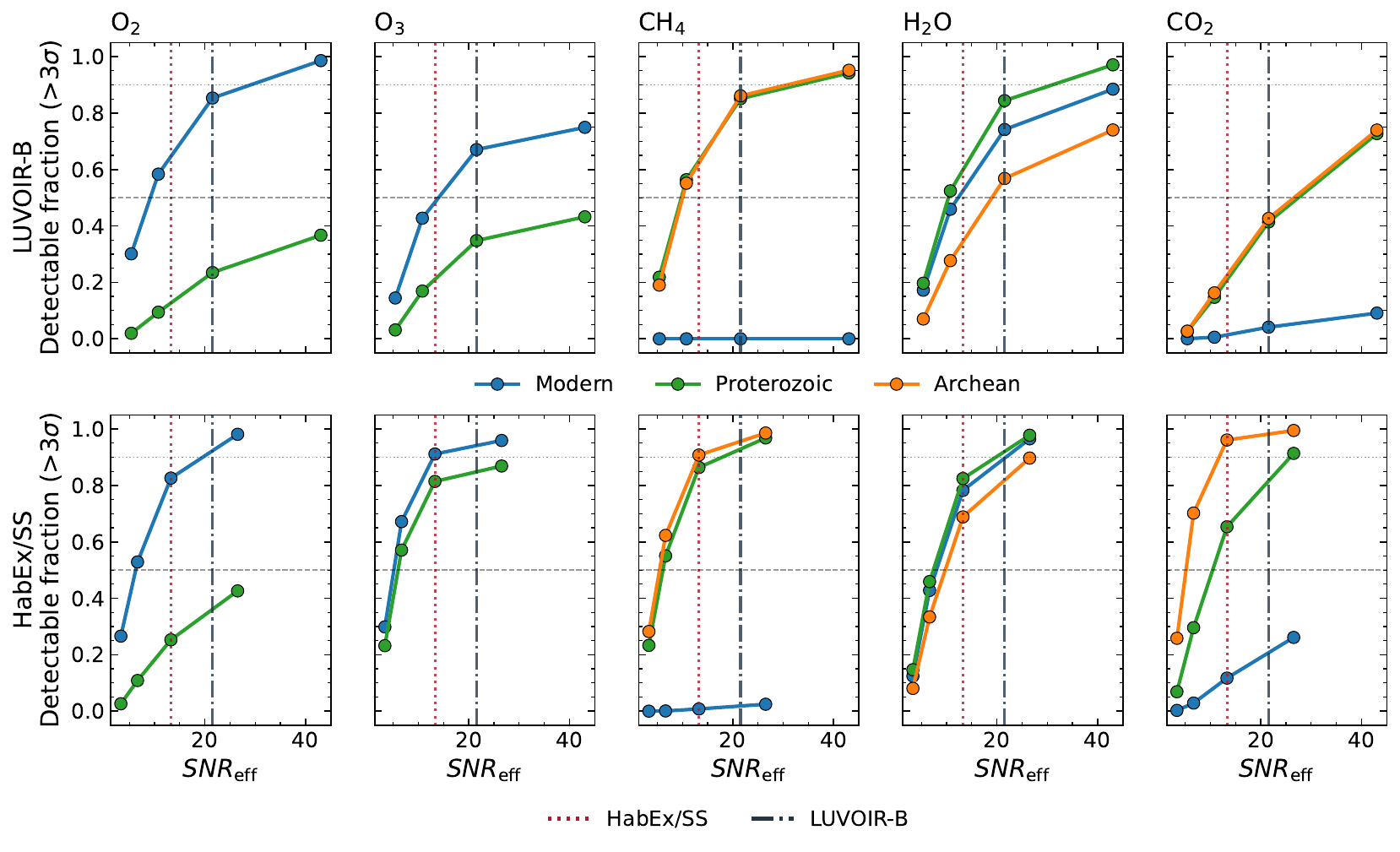}
\caption{Detectable fraction $f_{\mathrm{det}}$ at the $3\sigma$ physical-space criterion (equation~\ref{eq:phys_signif}) as a function of effective VIS-band SNR, $\text{SNR}_{\mathrm{eff}}$, evaluated on the held-out test set for the five chemically active species and the two instrument concepts (rows: LUVOIR-B, top; HabEx/SS, bottom). Curves are stratified by geological era, and the four points per curve correspond to noise scale factors $s \in \{4, 2, 1, 0.5\}$. Horizontal references at $f_{\mathrm{det}} = 0.5$ (dashed) and $0.9$ (dotted) mark the majority-detectable and high-confidence thresholds. Vertical lines indicate the median VIS-band $\text{SNR}_{\mathrm{eff}}$ at the default noise scaling $s=1$ for HabEx/SS (dotted, red) and LUVOIR-B (dash-dot, dark grey). Archean O$_2$ and O$_3$ are absent by construction (Table~\ref{tab:2}) and the corresponding curves are omitted.}
\label{fig:detection_fraction}
\end{figure*}

\subsubsection{Molecular detection thresholds}

The retrieval performance reported in Table~\ref{tab:model_metrics} and the variance decomposition characterize the network's accuracy in normalized model space, and the following analysis expresses these outputs as a $3\sigma$ detection criterion in mixing-ratio space for individual spectra. For this purpose, we propagate the combined uncertainty of equation~\ref{eq:sigma_total} back to mixing-ratio space via a first-order delta-method expansion of the generalized Hellinger transform adopted in our normalization scheme (Appendix~\ref{app:A}). For this monotonic power transform, the {mixing-ratio-space} detection significance simplifies to
\begin{equation}
    \mathcal{S}_j^{\mathrm{phys}} = \frac{\hat{x}_j}{\sigma_{x,j}} = \frac{\hat{z}_j}{n_j\,\sigma_{\mathrm{total},j}},
    \label{eq:phys_signif}
\end{equation}
where $\hat{z}_j$ is the bootstrap-mean prediction at noise scale $s$ and the factor $n_j$ in the denominator corrects for the differential compression introduced during normalization. A molecule is considered detected at $3\sigma$ when $\mathcal{S}_j^{\mathrm{phys}} > 3$, equivalently, when the credible interval $[\hat{x}_j - 3\sigma_{x,j},\,\hat{x}_j + 3\sigma_{x,j}]$ excludes zero abundance. From this criterion we define the detectable fraction $f_{\mathrm{det}}(E, j, s)$ as the proportion of test cases within geological era $E$ for which $z_{\mathrm{true},j} > 0$ and the above condition is satisfied at noise scale $s$. {Among the five species considered here, the exponent $n_j$ ranges from $n=2.75$ for O$_2$ to $n=15.70$ for O$_3$} (see Table \ref{tab:3}), and the correction therefore amplifies the physical-space uncertainty of the most strongly compressed species by nearly an order of magnitude, placing all species on a comparable detection footing. Molecular nitrogen, with no electric-dipole-allowed transitions in the UV-VIS-NIR range, would receive an artificially inflated significance under this framework and is therefore omitted from Fig. \ref{fig:detection_fraction}.

Fig. \ref{fig:detection_fraction} shows $f_{\mathrm{det}}$ as a function of effective VIS-band SNR for the five chemically active species, separated by geological era and instrument concept. Vertical reference lines mark the median VIS-band SNR achieved at the default noise scaling $s=1$ for each mission. {At the respective reference SNR, the signal best detected in oxygenated atmospheres is Modern O$_2$, reaching $f_{\mathrm{det}} = 0.85$ for LUVOIR-B and $0.83$ for HabEx/SS. Modern O$_3$ follows, and the two instruments separate sharply here: $f_{\mathrm{det}} = 0.91$ for HabEx/SS against $0.67$ for LUVOIR-B, with the same ordering in the Proterozoic ($0.81$ against $0.35$). This is the largest instrument difference anywhere in the figure. Because the O$_3$ variance is dominated by the MC-Dropout component, the difference arises mainly from the two networks and only secondarily from the spectra, as discussed below; the O$_3$ result is achieved despite the $n=15.70$ correction, owing to the spectral isolation of the Hartley-Huggins band identified in Section \ref{sec:4.2}. Proterozoic O$_2$, with mixing ratios approximately $10^{-2}$ times the present atmospheric level, lies below the $0.5$ majority-detectable threshold for both missions even at the highest SNR explored, reaching only $0.37$ and $0.43$ at $s=0.5$.} Methane displays the largest inter-era spread of any species: {Archean CH$_4$ at mixing ratios near $10^{-2}$ reaches $f_{\mathrm{det}} = 0.86$ and $0.91$ for LUVOIR-B and HabEx/SS at the default noise scale, whereas Modern CH$_4$ is the single most difficult case in the entire figure, at $f_{\mathrm{det}} = 0.000$ and $0.008$ and never exceeding $0.025$ at any SNR explored. This is consistent with the Modern-era performance collapse to $R^2 = 0.32$ and $0.35$ reported in Fig. \ref{fig:predtrue}.} Carbon dioxide follows the same abundance-driven pattern, rising from $0.04$ and $0.12$ in the Modern era to $0.43$ and $0.96$ in the Archean. Water vapor behaves differently: it is recovered above the $0.5$ threshold in every era and for both instruments, with values ranging from $0.57$ to $0.84$ and the best performance occurring in the Proterozoic. After Modern CH$_4$, the most challenging configurations are therefore Modern CO$_2$ and Proterozoic O$_2$, while H$_2$O remains comparatively well recovered.

\begin{figure*}
\centering
\includegraphics[width=\textwidth]{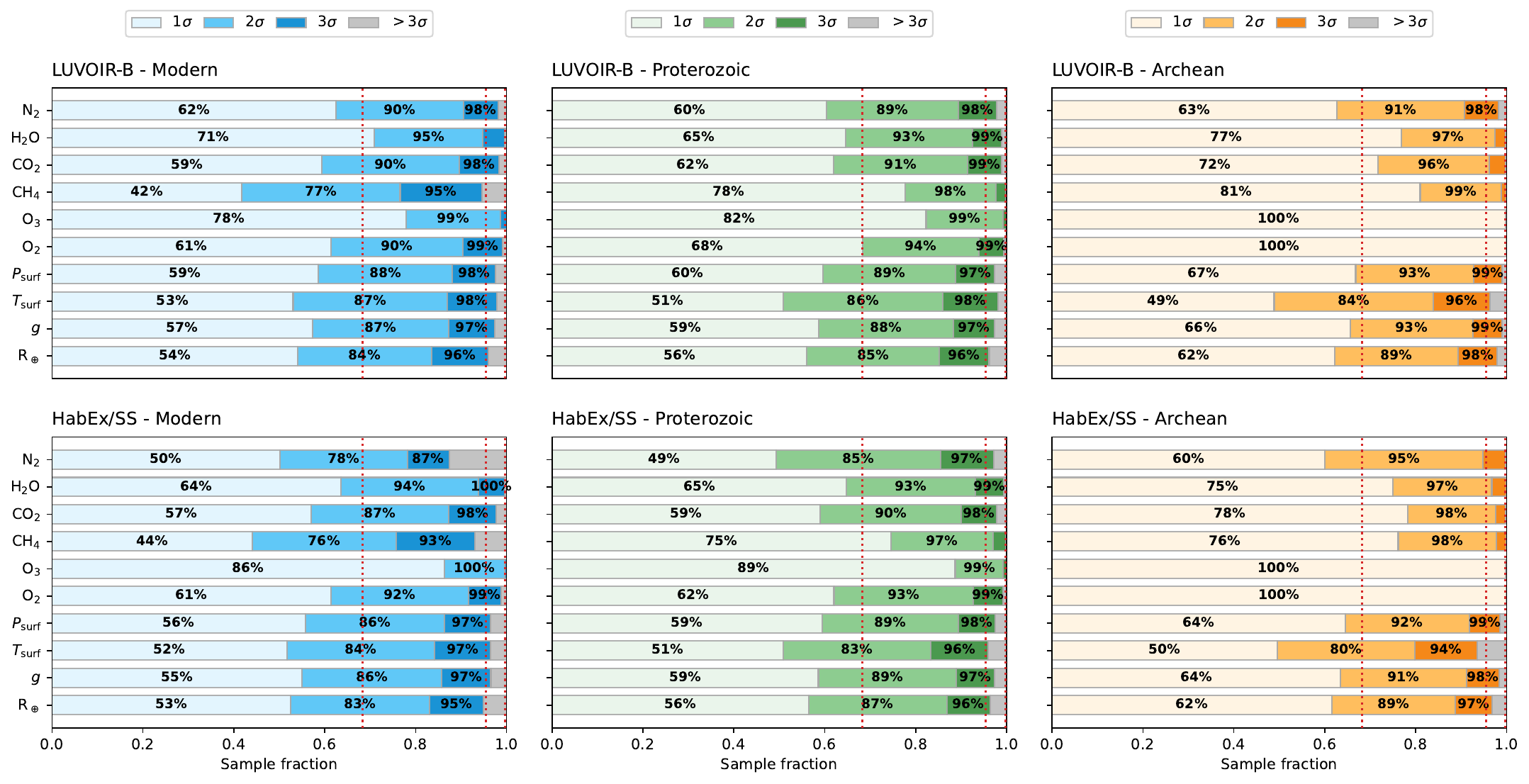}
\caption{Coverage calibration of the combined uncertainty $\sigma_{\rm total}$ across all retrieved parameters, stratified by instrument (rows) and geological era (columns). Each horizontal bar shows the cumulative fraction of test spectra falling within 1$\sigma$, 2$\sigma$, 3$\sigma$, and beyond 3$\sigma$ (gray) of the MC Dropout point estimate. Red dotted vertical lines indicate the Gaussian reference values.}
\label{fig:coverage}
\end{figure*}

The channel-level comparison accounts for part of this ordering. In the NIR, HabEx/SS has both higher detectable fractions for CH$_4$ and CO$_2$ and, most prominently, for Archean CO$_2$. Atmospheric abundance nevertheless remains the principal control within each instrument: Modern CH$_4$ has the highest NIR continuum of the three eras and still the lowest detectable fraction of any configuration, because its trace abundance dominates. The O$_3$ difference between the two concepts has a different origin, visible directly in the uncertainty decomposition. The median $\sigma_{\rm model}$ for O$_3$ is $0.0046$ for HabEx/SS against $0.0063$ for LUVOIR-B, and since $89\%$ to $93\%$ of the O$_3$ variance is epistemic (Fig. \ref{fig:uncerfra}), that difference propagates almost unchanged into $\sigma_{\rm total}$ and hence into $\mathcal{S}_j^{\mathrm{phys}}$.

The detection curves and the variance decomposition together define distinct observing priorities for O$_3$ and O$_2$. For O$_3$, the data fraction is approximately $7\%$ for LUVOIR-B and $11\%$ for HabEx/SS, leaving $93\%$ and $89\%$ of the quadrature variance in the MC-Dropout component. Halving the observational noise increases the Modern and Proterozoic O$_3$ detectable fractions by approximately $0.05$-$0.08$, concentrating the largest future gains in the inference model, training-set coverage, and uncertainty calibration. O$_2$ retains greater observational-noise leverage, especially in the Proterozoic. Halving the noise raises its detectable fraction from $0.24$ to $0.37$ for LUVOIR-B and from $0.25$ to $0.43$ for HabEx/SS; both values remain below the majority-detectable threshold.

\subsubsection{Coverage calibration}

Fig. \ref{fig:coverage} evaluates the frequentist coverage of the combined uncertainty estimate by reporting the cumulative fraction of test spectra whose true parameter values fall within the nominal 1$\sigma$, 2$\sigma$, and 3$\sigma$ intervals, stratified by geological era and instrument.

Across all eras and both telescopes, the 1$\sigma$ coverage is systematically below the Gaussian reference value of 68.3\% for most parameters. {Of the 56 informative parameter-era-instrument combinations, 43 fall below this reference, and the empirical coverage spans $42$-$89\%$.} For LUVOIR-B in the Modern era, for example, planetary radius reaches only {$54\%$} coverage within 1$\sigma$, and gravity {$57\%$}, while {H$_2$O lies closest to the Gaussian reference, at $71\%$.} {Surface temperature gives the most persistent undercoverage, remaining between $48.9$ and $53.0\%$ in every era and both instruments. Modern CH$_4$ gives the lowest individual values, at $42\%$ for LUVOIR-B and $44\%$ for HabEx/SS, matching the weak Modern-era retrieval performance reported in Section~\ref{sec:4}.}

{O$_3$ lies on the conservative side in every oxygenated subset, reaching $78$ and $86\%$ in the Modern era and $82$ and $89\%$ in the Proterozoic for LUVOIR-B and HabEx/SS, respectively. Its nominal 1$\sigma$ intervals are therefore conservative in all four cases. The instrument-level variance decomposition identifies the origin of the interval budget: observational noise contributes only $7\%$ and $11\%$ of the total O$_3$ variance for LUVOIR-B and HabEx/SS, leaving the interval width governed by the MC-Dropout component. The empirical coverage shows that the resulting uncertainty scale exceeds the central residual dispersion in each oxygenated subset.}

The predominance of 1$\sigma$ undercoverage across the remaining parameters identifies overconfidence in the combined estimator, consistent with documented calibration limitations of MC Dropout \citep{Folgoc2021}. We therefore use the nominal intervals as comparative uncertainty scales for ranking parameters, geological eras, and instrument concepts.

At the same time, the coverage improves at wider intervals. The 2$\sigma$ fractions span $76$-$100\%$, with 44 of the 56 informative combinations below the Gaussian reference value of $95\%$. Modern CH$_4$ gives the lowest values, followed by Modern N$_2$ in HabEx/SS. At 3$\sigma$, coverage exceeds $90\%$ in 55 of the 56 combinations; the sole exception is Modern N$_2$ in HabEx/SS, at $87\%$. The wider intervals therefore capture the bulk of the residual distribution. The fraction of outliers beyond 3$\sigma$ remains below $8\%$ for every chemical configuration except Modern N$_2$ in HabEx/SS, where it reaches $13\%$. Although $T_{\rm surf}$ has the most persistent undercoverage at 1$\sigma$, its 3$\sigma$ coverage remains between $94$ and $98\%$. This localizes its calibration error to the width of the central interval, while its residual tails remain controlled at 3$\sigma$.

In the Archean era, O$_2$ and O$_3$ both achieve essentially complete coverage within 1$\sigma$. These boundary-case values arise from the near-zero predictions and residuals produced when the network drives both outputs to zero throughout the anoxic subset, as documented in Section~\ref{sec:4}; they characterize the behavior of the imposed zero-abundance class.

Fig. \ref{fig:coverage} therefore defines the calibration domain of the uncertainty framework. Its strongest current application is comparative: identifying robust and fragile retrieval regimes, ranking parameter recoverability, and comparing observing architectures and atmospheric scenarios.

\section{Limitations}\label{sec:5}

The present results rely on a set of simplifying assumptions that define the scope of this work and provide the context for interpreting the retrieved parameters. The training data adopt isoabundant mixing-ratio profiles, obtained by replacing each vertically resolved template abundance with its pressure-weighted column mean (Section~\ref{sec:mixing_ratio}), while the P-T structure retains the vertical shape of the corresponding geological template. Composition is therefore decoupled from the P-T structure, which reduces forward-model complexity and isolates the retrieval signal to composition and bulk planetary properties, at the cost of discarding any information carried by vertical abundance gradients. This choice is well motivated for reflected-light spectra at the low spectral resolutions considered here, where sensitivity to vertical structure is inherently limited. It does not by itself explain the poor recovery of surface temperature, which arises instead because reflected light constrains the thermal structure only indirectly and because the liquid-water criterion compresses the intrinsic range of $T_{\rm surf}$ in the sample. The shape of this range introduces an additional sampling limitation. Although $T_{\rm surf}$ is calculated from instellation, Bond albedo, and infrared emissivity, $\varepsilon_{\rm IR}$ is drawn independently of atmospheric composition. The sample therefore does not encode the carbonate--silicate feedback expected to anticorrelate $p_{\rm CO_2}$ with instellation on Earth-like planets \citep{walker1981,kasting1993habitable,kopparapu2013habitable,lehmer2020}. Combined with the liquid-water selection, this leaves warm climates comparatively underrepresented and may reinforce the regression toward the center of the temperature distribution. Climate-coupled sampling will be considered in future work. 

The Archean template additionally omits the continuum opacity of organic hazes, whose radiative effects are degenerate with other parameters at these resolutions and whose inclusion we defer to future work alongside more realistic P-T profiles. A further boundary condition is that the training set is restricted to F- and G-type host stars, since K-type configurations did not pass the SNR threshold defined in Section \ref{sec:sample_selection}; extending the sample to K dwarfs, which are priority HWO targets \citep{cuntz2016exobiology}, will require either relaxed acceptance criteria or longer simulated integration times. The omission of clouds and hazes from all epoch templates is a related limitation of potentially broader impact. Heavily clouded atmospheres alter the continuum slope and molecular-feature contrast in ways that are not represented in the current training set. Depending on cloud-top pressure and optical depth, clouds can either mute absorption features or enhance their apparent contrast by increasing the reflected continuum outside the bands when sufficient absorbing gas remains above the cloud deck. Predictions for such targets cannot be considered reliable without retraining on clouded forward models. More generally, all conclusions in this work are conditional on the adopted forward-model and observing assumptions, including the PSG noise prescription, the benchmark exposure time, and the sample-selection criteria used to define viable targets.

As with any supervised ML retrieval model, predictive performance has been evaluated only within the parameter space spanned by the training data (Tables \ref{tab:1} and \ref{tab:2}). Spectra from atmospheric regimes not represented in this dataset will produce predictions whose reliability cannot be assessed without retraining. {More specifically, the modeled population is restricted to temperate rocky planets with $R_{\rm p}\leq1.23\,R_\oplus$ that satisfy the adopted surface-liquid-water criterion and whose compositions are generated around three N$_2$-balanced Earth-through-time templates. It therefore excludes, among other cases, hot and high-pressure CO$_2$-dominated Venus analogues, volatile-rich sub-Neptunes with H$_2$/He envelopes, frozen rocky planets, and atmospheres with bulk compositions far outside the adopted geological families. These regimes may introduce degeneracies in mean molecular weight, scale height, pressure, continuum opacity, and molecular-band contrast that are not represented in the present training set.} This is not a shortcoming unique to the present approach. Bayesian retrieval frameworks face the analogous constraint that posteriors are only meaningful within the prior volume. A practical distinction, however, is that the CNN does not \textit{natively} signal when an input falls outside its training distribution. Adding a dedicated out-of-distribution diagnostic is therefore identified as a priority direction for future work, as it would substantially strengthen the framework's usefulness for HWO target screening and triage. Similarly, because the fixed input dimension ties each model to a specific wavelength grid, direct application to instruments with different bandpass or spectral resolution would require retraining, although more flexible sequence-based architectures may alleviate this constraint in future work.

It is important to acknowledge that the high recoverability of both planetary radius ($R_p$) and surface pressure ($P_\text{surf}$) is partly driven by the data-generation procedure and does not solely reflect the spectroscopic information available in the spectra. Because $P_{\rm surf}$ {is obtained from $R_{\rm p}$ through the hydrostatic scaling relation of Section \ref{sec:2.2}, the only scatter being that of the sampled mass-radius exponent, the two quantities are almost perfectly coupled in the training set: their logarithms correlate at $r=0.999$, with a residual dispersion of only $1.9$\% in $P_{\rm surf}$ at fixed radius}. The CNN exploits this physical correlation in the training data to partially break the degeneracy between the two parameters. In practice, however, the HWO is not expected to provide such tight independent constraints on planetary radius from reflected-light spectra alone.

A final limitation concerns the uncertainty quantification framework. The combined uncertainty estimate based on bootstrap sampling and MC Dropout is informative for comparative analyses and relative trends, but it is not fully equivalent to a rigorously calibrated Bayesian posterior. The coverage analysis in Section \ref{sec:uncertainty} shows that the nominal $1\sigma$ intervals are too narrow for most parameters, so the framework is best suited for rapid screening and sensitivity analysis, while exact target-level credible intervals require more rigorous calibration. In addition, exact-zero abundance cases represent a special boundary condition in the label transformation, and small positive outputs near zero require caution, as they do not automatically constitute detections (Section \ref{sec:4}). These limitations do not invalidate the relative trends and comparative conclusions identified throughout this work, but they do define the level of rigor at which the results should be applied.

\section{Conclusions} \label{sec:conclusion}

We have presented a 1D CNN framework for the simultaneous retrieval of atmospheric composition and planetary properties from low-resolution reflected-light spectra of Earth-like exoplanets under HWO-like observing conditions. Trained on {109,945} synthetic PSG spectra spanning Archean, Proterozoic, and Modern atmospheric states for LUVOIR-B and HabEx/SS, the network functions both as a retrieval engine and as a diagnostic tool for evaluating what reflected-light characterization can recover from temperate rocky planets, and which aspects of the observing architecture matter most for the resulting science return.

The principal result is that HWO-like reflected-light observations appear capable, in principle, of recovering several key atmospheric tracers and bulk planetary properties with useful fidelity across multiple geological epochs. Ozone is the most robustly retrieved species, owing to the spectrally isolated Hartley-Huggins UV band, while O$_2$ is well constrained in oxygen-rich Modern-like atmospheres but degrades in the Proterozoic as its absorption weakens. H$_2$O and CH$_4$ are likewise recovered reliably when their spectral signatures are prominent. Among the physical parameters, radius, gravity, and surface pressure retain measurable information content, whereas surface temperature remains the clearest weak point of the reflected-light setup; the strong performance for $R_p$ and $P_{\mathrm{surf}}$ is partly aided by the {near-deterministic} correlation imposed during data generation, a constraint that is likely to be less favorable in real HWO observations.

At the task-aggregated level, the attribution density declines beyond approximately $1.7\,\mu$m. Its agreement with independent forward-model and Bayesian studies placing the relevant HWO boundary near $1.68$-$1.7\,\mu$m supports this wavelength as a mission-relevant trade point, although the present attribution analysis does not establish that longer wavelengths are dispensable for atmospheric regimes absent from the training population.

The integrated-gradient analysis provides an independent check that the network exploits physically meaningful structure. Its attribution maxima, read per detector channel, fall within one or two spectral bins of the O$_2$ A band at $762$ nm, and the CH$_4$ and CO$_2$ NIR bands at $1659$ and $1568$ nm, in agreement with the central wavelengths adopted in HWO spectroscopic trade studies, without any of that information having been supplied during training. In anoxic Archean atmospheres, where the gas is absent, the largest absolute O$_2$ attribution still lies near the A band, although with a much smaller amplitude.

Within the approximate uncertainty framework adopted here, the balance between the model-associated component and the resampled observational-noise contribution differs between the two concepts. For LUVOIR-B the budget is mixed, with six of the ten parameters model-dominated at the default PSG noise level; for HabEx/SS eight of the ten are dominated by observational noise, reaching $70$\% of the total variance for surface temperature. Only O$_3$ and CH$_4$ are model-dominated in both. Reducing observational noise therefore has broad leverage for HabEx/SS and parameter-dependent leverage for LUVOIR-B. Under our nominal $3\sigma$ criterion in mixing-ratio space, more than half of the Modern O$_3$ test cases meet the criterion at nominal noise for both instrument concepts, while fewer than half of the Proterozoic O$_2$ cases do so even at the lowest noise scaling explored. {Within the present trained networks and uncertainty framework, Modern-Earth CH$_4$ remains below the $3\sigma$ criterion even in the formal limit of vanishing observational noise, although this result is specific to the observing and inference framework adopted here.} The results therefore motivate improvements to the inference framework for model-dominated cases while showing that reductions in observational noise may remain beneficial for parameters in mixed or noise-sensitive regimes.

The comparison between instruments reveals no architecture that is uniformly superior. HabEx/SS attains higher detectable fractions for O$_3$ and for the NIR species CH$_4$ and CO$_2$, whereas LUVOIR-B recovers surface temperature more accurately; these differences must be read alongside the fact that the two networks were optimized independently. The Archean tests show that both concepts are resilient against false biosignature claims in truly anoxic atmospheres: an exactly zero abundance is returned for $99.8$ to $100$\% of the anoxic spectra, and the residual false-positive rate stays below $0.21$\% for O$_3$ and $0.03$\% for O$_2$, with the few non-zero values lying at the floor of the single-precision representation.

These conclusions are conditional on a set of simplifying assumptions, discussed fully in Section~\ref{sec:5}. The present pipeline therefore serves as a rapid and informative tool for HWO retrieval analysis, while fully calibrated posterior inference remains necessary for detailed characterization of individual planets. Beyond demonstrating the retrieval framework, our results identify which atmospheric quantities are realistically accessible to reflected-light observations, which spectral regions provide the relevant diagnostic information, and where the main limitations on HWO biosignature retrieval arise.

\section*{Acknowledgements}

We gratefully acknowledge the financial support from the Brazilian agency CAPES (grant No. 88887.622098/2021-00), as well as the STELLAR TEAM at the Federal University of Ceará for our discussions and collaborative support. Part of the research was carried out at the Jet Propulsion Laboratory, California Institute of Technology, under a contract with the National Aeronautics and Space Administration (80NM0018D0004). We also extend our thanks to Geronimo Villanueva for his assistance with setting up the PSG, and to Yui Kawashima for providing data on the Proterozoic Earth.

\section*{Data Availability}

The trained CNN models for both the HabEx/SS and LUVOIR-B configurations, as well as the corresponding training, validation, and test datasets, have been deposited on Zenodo (DOI: \href{https://doi.org/10.5281/zenodo.21959726}{10.5281/zenodo.21959726}).

\section*{Conflict of Interest}

Authors declare no conflict of interest.



\bibliographystyle
{rasti}
\bibliography{example} 




\appendix

\section{Dataset structure} \label{app:dataset_structure}

For each simulated planet, the network receives three one-dimensional spectral arrays corresponding to the UV, VIS, and NIR bands defined in Section~\ref{sec:telescope_properties}, drawn from the noisy apparent-albedo spectrum $A_{\rm noisy}(\lambda)$ (equation \ref{eq:noisy_spectrum}). The target variables are organized into three groups matching the multi-head output architecture described in Section \ref{sec:3}:

\begin{enumerate}
\item Physical parameters: planetary radius, surface gravity, surface temperature, and surface pressure.
\item Primary biosignatures: O$_2$ and O$_3$, treated as a separate group due to their strong diagnostic features in the UV/VIS range and their central role as indicators of biological oxygenation in Earth-like atmospheres.
\item Other chemical abundances: all remaining atmospheric species (e.g., H$_2$O, CO$_2$, CH$_4$, N$_2$).
\end{enumerate}

This hierarchical target structure is identical for both the LUVOIR-B and HabEx/SS models. {After all physical-selection and signal-to-noise cuts, the 109,945 accepted planet realizations were stratified by geological epoch and divided into training (80\%; 87,954 realizations), validation (10\%; 10,993), and held-out test (10\%; 10,998) subsets. Within each epoch, realizations were shuffled using a fixed random seed and assigned to the three subsets in these proportions. The splitting unit was the complete planet realization, which contains the paired LUVOIR-B and HabEx/SS spectra; thus, spectra of the same planet could not occur in different subsets. The test set comprises 3,353 Archean, 3,701 Proterozoic, and 3,944 Modern realizations. The validation subset was used for hyperparameter optimization, early stopping, and model selection, whereas the test subset was not used during any stage of model fitting or selection. Normalization statistics were calculated exclusively from the training subset and then applied unchanged to the validation and test subsets, as described in Appendix \ref{app:A}.}

{As a quantitative check that the held-out sample represents the same population as the training data, we compared the distributions of all ten target variables. The largest two-sample Kolmogorov-Smirnov statistic was 0.014 over the complete data set and 0.027 when the comparison was performed separately within each geological epoch; the corresponding largest standardized mean differences were 0.024 and 0.043, respectively. These small differences confirm that the training and test subsets have statistically similar parameter distributions.} Full implementation details are documented in the public repository at \url{https://github.com/SarahBarbosa/geexhp}.

\section{Data normalization}\label{app:A}

Neural network training requires that input features and target variables share comparable numerical ranges. Without normalization, features with large ranges can dominate gradient updates and hinder convergence, while different target distributions can bias optimization. To address this, we applied three scaling approaches, each designed to match the statistical characteristics of: (i) the input spectra, (ii) the physical parameters and (iii) the chemical abundances. In all cases, the required statistics (e.g., mean, variance, extrema) were computed \textit{only} on the training set and subsequently applied to the validation and test splits to prevent data leakage.

For every input spectrum, the noisy apparent albedo $A_{\text{noisy}}(\lambda_j)$ is standardized (Z-score normalized) {per instrument channel}:
\begin{equation}\label{a1}
    \overline{A}_{\text{noisy}}(\lambda_j) = \frac{A_{\text{noisy}}(\lambda_j) - \mu_c}{\sigma_c},
\end{equation}
{where $c$ denotes the channel containing bin $\lambda_j$ (UV, VIS or NIR), and $\mu_c$ and $\sigma_c$ are the mean and standard deviation over all bins of that channel and all training spectra.}

The physical parameters are rescaled using a Min-Max transformation to map values into the $[0, 1]$ interval:
\begin{equation}\label{eq:minmax}
    \overline{y}_{ij} = \frac{y_{ij} - \min_j(y_{ij})}{\max_j(y_{ij}) - \min_j(y_{ij})},
\end{equation}
where $y_{ij}$ is the raw value of variable $i$ for sample $j$, and the extrema ($\min, \max$) are computed over the entire training dataset.

The chemical targets (biosignatures and other species) span several orders of magnitude and contain true zeros (e.g., in the Archean era). Standard logarithmic scaling is therefore unsuitable, as $\log(0)$ is undefined and the large range causes high abundances to cluster near $\log(1)=0$. 

To mitigate this sparsity and zero-inflation, we adopted a power‑law mapping inspired by the \textit{Hellinger transform} \citep{legendre2001ecologically, CONDE2018161}. For each sample $j$, abundances are first converted into per-sample proportions,
\begin{equation}
p_{ij}=\frac{y_{ij}}{\sum_{i'} y_{i'j}},
\end{equation}
where $y_{ij}\ge 0$ denotes the raw abundance of species $i$ in sample $j$ and the sum runs over all modeled gases. We then apply a species-dependent power transform,
\begin{equation}\label{eq:hellinger_general}
\overline{H}_{ij} = \left(p_{ij}\right)^{1/n_i}, \quad n_i>0,
\end{equation}
where $n_i$ controls the compression/stretching of the distribution for species $i$ (note that $n_i=2$ recovers the standard Hellinger transform and $n_i=1$ leaves $p_{ij}$ unchanged).

For each species we scan $n_i \in [0.05,50]$ and select the value that maximizes the width of the interval occupied by the transformed distribution, measured on a 200-bin histogram. The adopted values are listed in Table \ref{tab:3}, and the selection procedure is illustrated in Fig. \ref{fig:2}.

\begin{table}
    \small
    \centering
    \setlength{\tabcolsep}{4pt}
    \caption{Selected exponent $n_i$ for each transformed target in equation~\ref{eq:hellinger_general}.}
    \label{tab:3}
\begin{tabular}{lccccccc}
    \hline
    {Parameter} & CH$_4$ & CO$_2$ & H$_2$O & N$_2$ & O$_2$ & O$_3$ \\
    \hline
    {Best $n_i$} & {8.65} & {5.50} & {5.80} & {0.15} & {2.75} & {15.70} \\
    \hline
\end{tabular}
\end{table}

\begin{figure}
\centering
\includegraphics[width=\columnwidth]{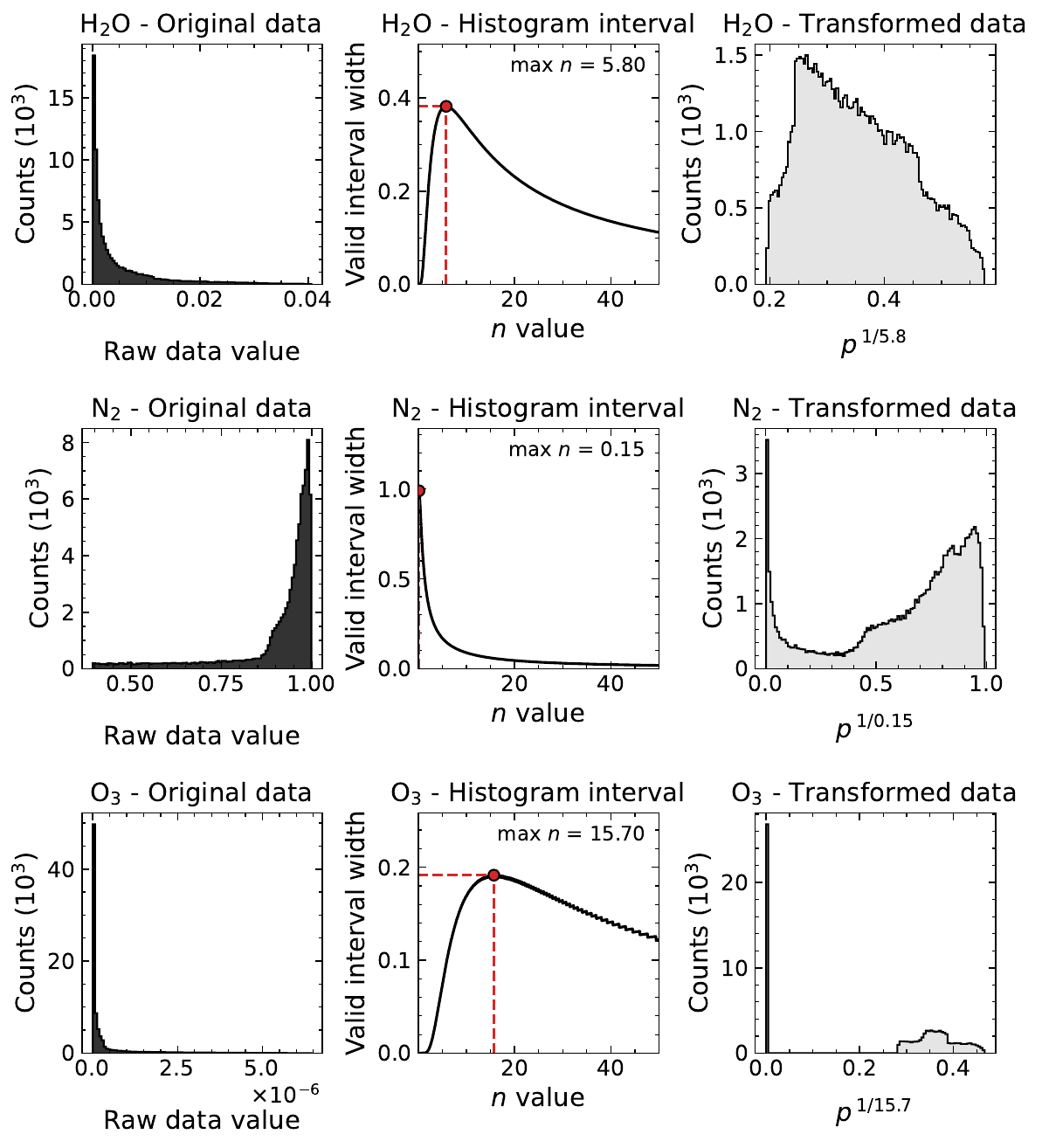}
\caption{Examples of species distributions. The left column shows the original distribution of the data. The middle column shows the interval analysis of the histogram and the best value of $n_i$ indicated. The right column shows the new distribution of each species after the transformation in equation \ref{eq:hellinger_general} is applied with the best $n_i$.}
\label{fig:2}
\end{figure}

\section{Evaluation metrics}\label{app:metrics_explanation}

The five metrics presented in Table \ref{tab:model_metrics} are complementary as they capture different aspects of the model's accuracy.

Let $\{(y_i,\hat{y}_i)\}_{i=1}^{N}$ denote the true and predicted values for a given parameter, and let $\bar{y}=(1/N)\sum_{i=1}^{N}y_i$ be the sample mean. The coefficient of determination, $R^2$, is given by
\begin{equation}
R^2 = 1 - \dfrac{\sum_{i=1}^{N}(y_i-\hat{y}_i)^2}{\sum_{i=1}^{N}(y_i-\bar{y})^2}.
\end{equation}
It quantifies the fraction of variance in $y$ explained by the predictions \citep{Pedregosa2011}. An $R^2=1$ indicates perfect agreement, $R^2=0$ corresponds to performance equivalent to predicting the constant mean $\hat{y}_i=\bar{y}$, and $R^2<0$ occurs when prediction errors exceed those of the mean predictor.

To quantify deviations in the transformed model-output space, we compute the mean absolute error (MAE) and root-mean-square error (RMSE):
\begin{equation}
\mathrm{MAE} = \frac{1}{N} \sum_{i=1}^{N} |\hat{y}_i - y_i|,
\qquad
\mathrm{RMSE} = \sqrt{\frac{1}{N} \sum_{i=1}^{N} (\hat{y}_i - y_i)^2}.
\end{equation}
MAE measures the average magnitude of the residuals and treats all deviations linearly, which makes it comparatively robust to occasional large outliers \citep{Willmott2005}. For this reason, MAE was not only adopted as an evaluation metric but also implemented as the loss function during model training. RMSE, in contrast, weights errors quadratically and therefore amplifies larger discrepancies. By construction, $\mathrm{RMSE} \ge \mathrm{MAE}$, with equality only when all residuals have identical magnitude.

Because several predicted quantities span multiple orders of magnitude, we additionally report scale-independent percentage metrics. The mean absolute percentage error (MAPE) and weighted absolute percentage error (WAPE) are defined as
\begin{equation}
\text{MAPE} = \dfrac{100}{N} \sum_{i=1}^{N} \left| \frac{\hat{y}_i - y_i}{y_i} \right|,
\qquad
\text{WAPE} = 100 \times \dfrac{\sum_{i=1}^{N} |\hat{y}_i - y_i|}{\sum_{i=1}^{N} |y_i|}.
\end{equation}
MAPE expresses each error relative to its corresponding true value, enabling comparisons across parameters with different physical scales \citep{Hyndman2006}. However, it becomes unstable when $y_i$ approaches zero, as small absolute deviations can produce disproportionately large percentage errors \citep{Kim2016}. WAPE mitigates this limitation by normalizing the aggregate absolute error by the total absolute signal, thereby reducing sensitivity to near-zero denominators and providing a more stable summary statistic for skewed distributions \citep{Kolassa2007}.

\section{Projected integrated-gradient attribution spectra}\label{app:ig_projection}

To complement the two-dimensional IG heatmaps shown in Fig. \ref{fig:ig_heatmap}, we define a one-dimensional wavelength-priority metric obtained by projecting the attribution maps onto the wavelength axis. The goal of this procedure is to summarize, for a given instrument and geological era, which spectral {regions carry the most retrieval leverage} across the retrieval task as a whole.

Let $\mathrm{IG}_{n,k}(\lambda_j)$ denote the integrated gradient at wavelength bin $\lambda_j$ for test spectrum $n$ and retrieved output $k$. Because the sign of IG indicates whether increasing the flux in a given bin raises or lowers the prediction, whereas the present objective is to measure the overall influence of each bin independent of direction, we first take the absolute attribution,
\begin{equation}
A_{n,k}(\lambda_j) = \left|\mathrm{IG}_{n,k}(\lambda_j)\right|.
\end{equation}
We then normalize this attribution spectrum across wavelength for each sample-output pair,
\begin{equation}
\widetilde{A}_{n,k}(\lambda_j) = \frac{A_{n,k}(\lambda_j)} {\sum_{j'} A_{n,k}(\lambda_{j'})}, \label{eq:ig_norm_per_output}
\end{equation}
so that each valid output contributes unit total attribution when integrated over wavelength. This step is essential because without it outputs with intrinsically larger raw IG amplitudes would dominate the projected score, even if their attribution were concentrated in only a limited subset of the science problem. The normalization therefore makes the projection sensitive to the \emph{distribution} of attribution across wavelength, independent of the raw numerical scale of the IG values.

The projected wavelength-priority score {for era $e$} is then defined by averaging equation~(\ref{eq:ig_norm_per_output}) over all valid test spectra {of that era} and over the $K$ retrieved outputs,
\begin{equation}
S_{e}(\lambda_j) = \frac{1}{K} \sum_{k=1}^{K} \left\langle \widetilde{A}_{n,k}(\lambda_j)
\right\rangle_{n \in \mathcal{V}_k}, \label{eq:projected_ig_score}
\end{equation}
where $\mathcal{V}_k$ denotes the subset of samples for which output $k$ has non-negligible total attribution, defined as $\mathcal{V}_k=\{\,n:\sum_j A_{n,k}>\eta\,\mathrm{median}_{n'}\sum_j A_{n',k}\}$. {The threshold is defined relative to the median because the outputs are trained in power-transformed spaces with substantially different exponents (Appendix~\ref{app:A}), making their raw IG magnitudes not directly comparable. With $\eta=10^{-3}$ the criterion is inactive for the network presented here: every test spectrum is retained for all outputs, eras and instruments, with the sole exception of N$_2$ in the HabEx/SS Proterozoic and Archean subsets, where $0.4$ and $1.3$\% of spectra are removed. The valid fractions are identical for $\eta$ between $10^{-4}$ and $10^{-2}$}. By construction,
\begin{equation}
\sum_j S_{e}(\lambda_j) = 1,
\end{equation}
up to numerical precision.

{Equation~(\ref{eq:projected_ig_score}) is a per-\emph{bin} quantity. Plotted against wavelength, a per-bin score would therefore assign a wide UV channel more weight simply because it spans a broader interval, and the two instruments would not be directly comparable. We accordingly convert it to an attribution density,}
\begin{equation}
s_e(\lambda_j) = \frac{S_e(\lambda_j)}{\Delta \lambda_j},
\qquad
\int s_e(\lambda)\,\mathrm{d}\lambda = 1 ,
\label{eq:ig_density}
\end{equation}
{where $\Delta\lambda_j$ is the width of bin $j$. Fig. \ref{fig:ig_projected} shows $s_e(\lambda)$, such that the area under each curve measures the attribution associated with a given wavelength interval, independent of the individual bin widths. The two normalizations rank the channels differently, which is why Section \ref{sec:4.2} distinguishes them explicitly: per bin the largest single values of $S_e$ fall in the UV in four of the six instrument-era combinations, whereas per micron the visible carries the higher mean density in all six. The channel fractions quoted in Section \ref{sec:4.2} are computed from $S_e$ before this conversion, so that they sum to unity across the three channels. Dividing by $\Delta\lambda_j$ assumes that the network sensitivity assigned to each bin is uniformly distributed across its width, since integrated gradients operate on the binned spectrum and do not resolve the underlying continuum.}

\section{Uncertainty estimation approach}\label{app:uncertainty}

This appendix presents the details of the two-component uncertainty approach summarized in subsection \ref{sec:uncertainty}.

\subsection{Data uncertainty via spectral bootstrap}

The spectra used for training include a single pre-generated noisy realization per sample, produced by the Gaussian noise model defined in equation \eqref{eq:noisy_spectrum}. However, a bootstrap estimate of data uncertainty requires $B$ statistically independent noise realizations for the same underlying spectrum. If we were to start from $A_{\text{noisy}}(\lambda)$, we would unintentionally introduce a second, independent noise draw on top of the original one, thereby increasing the overall noise amplitude compared to the intended model. At the default noise level, this inflates the data uncertainty by a factor of $\sqrt{2}$.

To avoid this issue, what we actually do is recompute $A_{\text{noisy}}(\lambda)$ by introducing a scaling factor $s$:
\begin{equation}\label{eq:noisy_spectrum2}
\tilde{A}^{(b)}(\lambda) \sim \mathcal{N}(A(\lambda), s \sigma(\lambda)),
\end{equation}
so that each draw is statistically consistent with the noise distribution used to train the network. In fact, the scale factor $s$ provides an idealized rescaling of the complete wavelength-dependent PSG noise vector. The case $s=1$ reproduces the default noise prescription, $s=0.5$ uniformly halves the adopted noise vector, and $s>1$ represents progressively larger observational noise.

Each realization is Z-score normalized using the training-set mean and standard deviation {of the corresponding instrument channel} (equation \ref{a1}) and passed through the network with dropout disabled (\texttt{training=False}). For a given parameter $k$, the data uncertainty of spectrum $i$ is
\begin{equation}
\sigma_{\text{data}}^{(i,k)} = \text{std} \left\{f_k \left(\tilde{A}^{(b)}_i\right)
\right\}_{b=1}^{B},
\end{equation}
where $f_k$ denotes the network output for parameter $k$, evaluated deterministically.  


\subsection{Model uncertainty via Monte Carlo Dropout}

Unlike the bootstrap, MC Dropout takes a single fixed input through the network $N_{\rm MC}$ times, {resampling the stochastic dropout masks at each pass}. In practice, this is implemented by performing inference with dropout enabled (\texttt{training=True}). During each forward pass, an independent SpatialDropout1D mask is applied to the convolutional feature maps, and a separate Bernoulli mask is applied to the dense-layer activations. The resulting model uncertainty for parameter $k$ is
\begin{equation}
\sigma_{\mathrm{model}}^{(i,k)} = \text{std}\left\{ f_k^{(t)}\left(A_{i, \text{noisy}}\right) \right\}_{t=1}^{N_{\rm MC}},
\end{equation}
where $f_k^{(t)}$ denotes the network output under the $t$-th dropout mask. The point estimate $\hat{y}^{(i,k)}$ is taken as the mean over the $N_{\rm MC}$ passes, serving as the MC Dropout posterior mean.

\section{Per-sample predictive distributions}\label{app:cornerplot}

The aggregate coverage analysis presented in Section \ref{sec:uncertainty} collapses the joint predictive distribution into marginal fractions per parameter. To complement that population-level view, Figs. \ref{fig:corner-modern}, \ref{fig:corner-proterozoic}, and \ref{fig:corner-archean} present the full joint predictive distribution for representative Modern, Proterozoic, and Archean test spectra, respectively.

{Four ensembles are overlaid on the same axes, with colour encoding the instrument and line style the type of uncertainty: the spectral bootstrap for LUVOIR-B and HabEx/SS, each constructed from $5{,}000$ independent noise realizations, and the MC Dropout ensemble for each instrument, each constructed from $5{,}000$ stochastic forward passes. Only the LUVOIR-B bootstrap is filled; two filled distributions conceal one another wherever they overlap, and which one is hidden changes from era to era, so the HabEx/SS bootstrap is drawn as an outline. Contour levels enclose 39.3\% and 86.5\% of the probability mass, corresponding to the $1\sigma$ and $2\sigma$ credible regions of a 2D Gaussian.}

\begin{figure*}
\centering
\includegraphics[width=\textwidth]{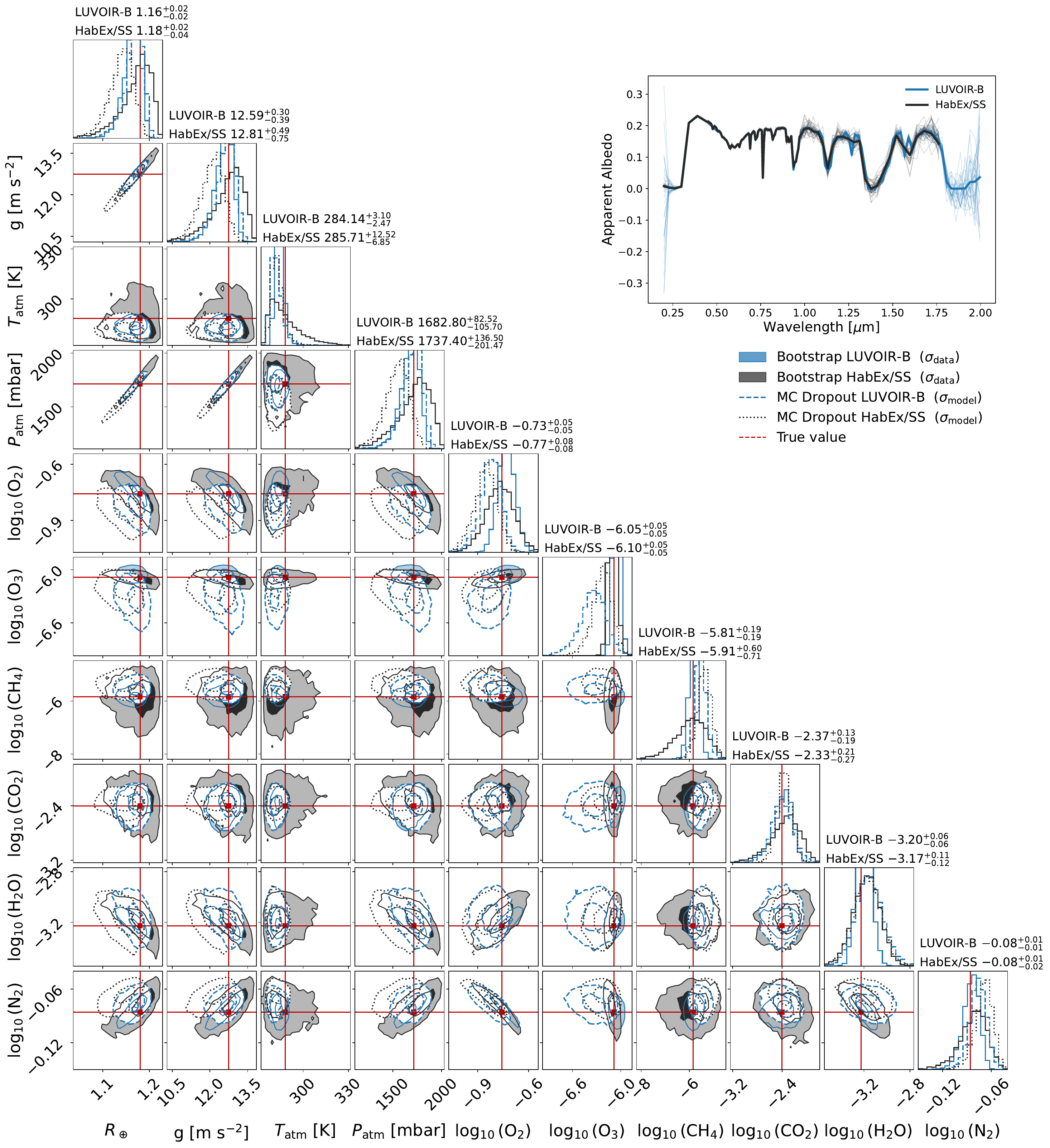}
\caption{Joint predictive {distributions} for a representative Modern-era test spectrum, projected onto all 45 pairwise parameter combinations. {Colour encodes the instrument and line style the type of uncertainty. Blue filled contours: spectral bootstrap for LUVOIR-B ($\sigma_{\rm data}$, $B = 5{,}000$); dark gray solid contours: spectral bootstrap for HabEx/SS ($\sigma_{\rm data}$, $B = 5{,}000$); blue dashed contours: MC Dropout for LUVOIR-B ($\sigma_{\rm model}$, $N_{\rm MC} = 5{,}000$); dark gray dotted contours: MC Dropout for HabEx/SS ($\sigma_{\rm model}$, $N_{\rm MC} = 5{,}000$). Only the LUVOIR-B bootstrap is filled, since two filled distributions conceal one another where they overlap. The LUVOIR-B ensembles are drawn in the colour of the corresponding geological era, blue for the Modern case shown here.} Contour levels enclose 39.3\% and 86.5\% of the probability mass ($1\sigma$ and $2\sigma$ for a 2D Gaussian). Red {solid} lines indicate the true parameter values. {The values above each diagonal panel are the median and $\pm1\sigma$ from the 16th, 50th, and 84th percentiles of the two bootstrap ensembles, labeled by instrument.} The inset shows the noiseless apparent-albedo spectrum for both instruments alongside 30 {noise realizations drawn for each of them at the observed SNR, which is the input distribution the bootstrap contours propagate}.}
\label{fig:corner-modern}
\end{figure*}

\begin{figure*}
\centering
\includegraphics[width=\textwidth]{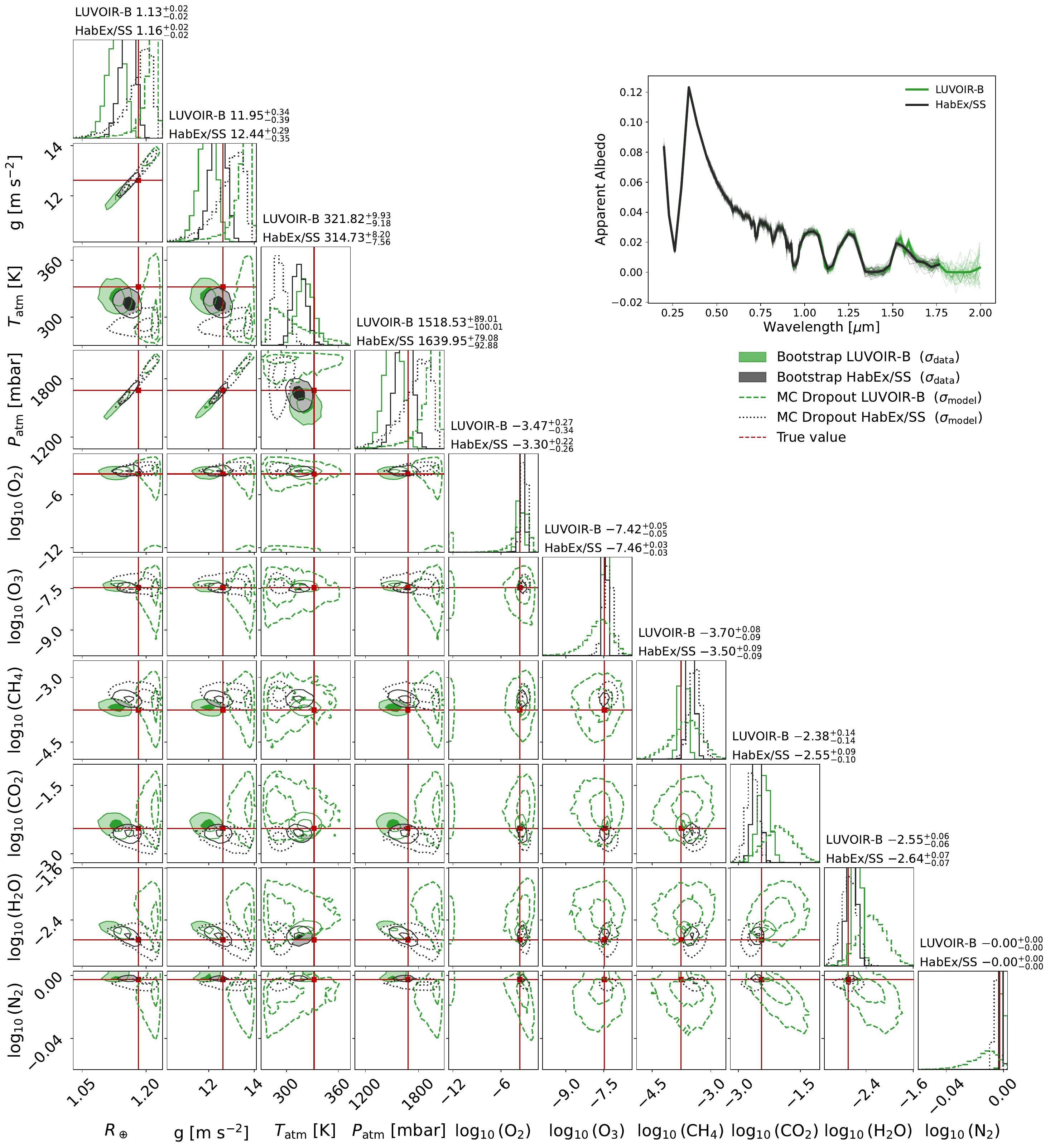}
\caption{The same as Fig. \ref{fig:corner-modern} but for a representative Proterozoic era test spectrum{, with the LUVOIR-B ensembles drawn in green, the color adopted for that era}.}
\label{fig:corner-proterozoic}
\end{figure*}

\begin{figure*}
\centering
\includegraphics[width=\textwidth]{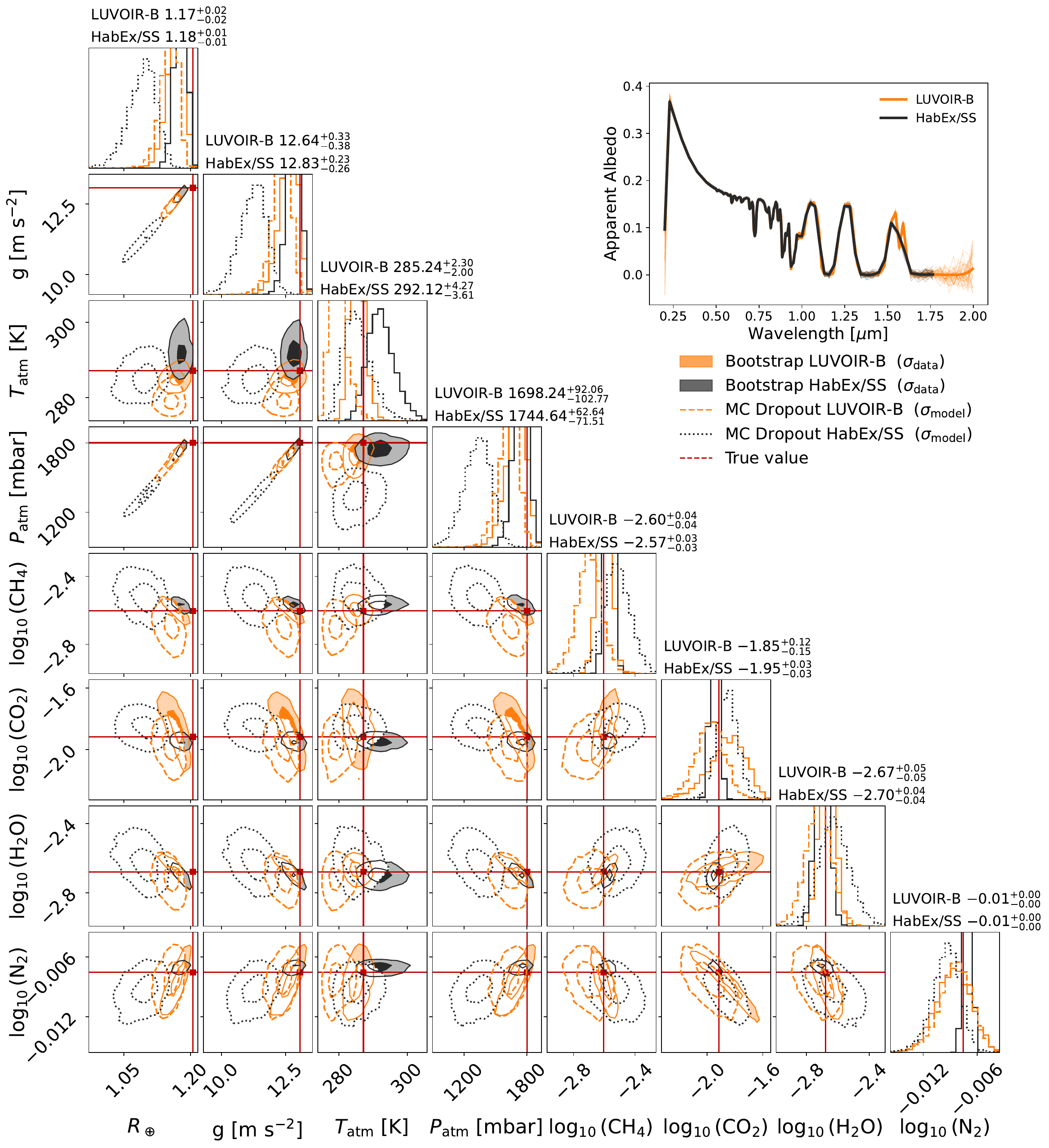}
\caption{The same as Fig. \ref{fig:corner-modern} but for a representative Archean era test spectrum{, with the LUVOIR-B ensembles drawn in orange, the color adopted for that era. This figure is eight-dimensional and shows 28 pairwise combinations, where O$_2$ and O$_3$ are identically zero in the Archean by construction (Table \ref{tab:2}), and a predictive distribution for an identically zero target is not meaningful}.}
\label{fig:corner-archean}
\end{figure*}


\bsp	
\label{lastpage}
\end{document}